%% file: sigir26.tex
\documentclass[sigconf]{acmart}

\usepackage{hyperref}
\usepackage{multirow}
\usepackage{subcaption}
\usepackage{caption}
\usepackage{tabularx}
\usepackage{setspace}
\usepackage{amsmath}
\usepackage[english]{babel}
\definecolor{mygray}{gray}{0.5}
\usepackage{float}
\usepackage{flushend}
\usepackage{mathtools}
\usepackage{soul}
\usepackage{arydshln}
\usepackage{algpseudocode}
\usepackage{algorithm}
\usepackage{url}
\usepackage{xcolor}

\usepackage{graphicx}

\usepackage{enumitem}

\long\def\comment#1{}
\long\def\comments#1{}

\copyrightyear{2026}
\acmYear{2026}
\setcopyright{cc}
\setcctype{by}
\acmConference[SIGIR '26]{Proceedings of the 49th International ACM SIGIR Conference on Research and Development in Information Retrieval}{July 20--24, 2026}{Melbourne, VIC, Australia}
\acmBooktitle{Proceedings of the 49th International ACM SIGIR Conference on Research and Development in Information Retrieval (SIGIR '26), July 20--24, 2026, Melbourne, VIC, Australia}
\acmDOI{10.1145/3805712.3809625}
\acmISBN{979-8-4007-2599-9/2026/07}

\ccsdesc[500]{Information systems~Learning to rank}
\keywords{Learned Sparse Representations; Sparse Retrieval; Sparsification; Efficiency}

\author{Wentai Xie}
\affiliation{%
  \institution{University of California, Santa Barbara}
  \city{Santa Barbara}
  \state{California}
  \postcode{93106}
  \country{USA}
}
\email{wentaixie@ucsb.edu}
\author{Parker Carlson}
\affiliation{%
  \institution{University of California, Santa Barbara}
  \city{Santa Barbara}
  \state{California}
  \postcode{93106}
  \country{USA}
}
\email{parker_carlson@ucsb.edu}
\author{Shanxiu He}
\affiliation{%
  \institution{University of California, Santa Barbara}
  \city{Santa Barbara}
  \state{California}
  \postcode{93106}
  \country{USA}
}
\email{shanxiuhe@ucsb.edu}
\author{Tao Yang}
\affiliation{%
  \institution{University of California, Santa Barbara}
  \city{Santa Barbara}
  \state{California}
  \postcode{93106}
  \country{USA}
}
\email{tyang@cs.ucsb.edu}
\renewcommand{\shortauthors}{Wentai Xie,  Parker Carlson, Shanxiu He, and Tao Yang}

\begin{document}

\title{
Adaptive Sparsity Optimization with Learnable Soft Top-$K$ and  Per-Term Thresholding
for Efficient Retrieval 
}

\begin{abstract}

\input{sigir26abs}

\end{abstract}

\maketitle

\input{sigir26intro}
\input{sigir26background}
\input{sigir26design}

\input{sigir26method}

\input{sigir26experiment}
\input{sigir26ablation}
\input{sigir26concl}

\balance
\newpage
\bibliographystyle{ACM-Reference-Format}
\normalsize
\bibliography{bib/2025.bib,bib/2024refer.bib,bib/thres.bib,bib/2022extra.bib,bib/2022refer.bib,bib/jinjin_thesis.bib,bib/mise.bib,bib/ranking.bib,bib/reference.bib,bib/url.bib,bib/distill.bib,bib/2026wentai.bib}

\end{document}

%% file: sigir26abs.tex
Recent work on neural sparse retrieval has demonstrated strong relevance by leveraging Large Language Models (LLMs) for semantic term expansion. However, learned models paired with previous sparsification techniques still yield overly long document and query
vectors partly due to a large LLM vocabulary, imposing a serious challenge to retrieval time and space efficiency. This paper proposes a scheme for optimizing model sparsity through a synergy of adaptive strategies, including learnable soft top-$K$, per-term thresholding, 
and FLOPs regularization to increase the sparsity of query and document vectors. Experimental results with Lion-SP model on the MS MARCO and BEIR datasets demonstrate that the proposed scheme can outperform the 
baselines by significantly reducing the average query and document lengths. Our scheme can achieve much shorter retrieval latency and lower storage cost while maintaining highly competitive relevance.

%% file: sigir26intro.tex
\section{Introduction}

Large-scale text search and  retrieval-augmented generation~\cite{2020NIPS-Lewis-RAG}
depend on efficient document retrieval to identify the top-$k$ relevant results for re-ranking or LLM-based inference. Learned sparse retrieval techniques~\cite{Zamani2018SNRM,
Dai2020deepct, Mallia2021deepimpact, Lin2021unicoil,2021NAACL-Gao-COIL,
Formal2021SPLADE, Formal2021SPLADEV2,Formal_etal_SIGIR2022_splade++}
based on BERT have demonstrated strong effectiveness in neural retrieval. These methods leverage transformer-based models to expand document tokens and assign learned weights, producing sparse representations with high relevance. 
At the same time, they can take advantage of traditional inverted index based retrieval techniques~\cite{Mallia2017VBMW,Mallia2021deepimpact} , enabling efficient retrieval on low-cost CPU platforms.

Recently, the Lion-SP model~\cite{zeng2025scalingsparsedenseretrieval}
represents a shift toward using large-scale decoder-only architectures for sparse vector generation with LLaMA-3~\cite{grattafiori2024llama3herdmodels}. 
Although Lion adopts the SPLADE methodology by using FLOPs regularization~\cite{Paria2020Minimizing, Formal2021SPLADE,lassance2022efficiency} 
 to minimize non-zero weights, Lion-SP still generates vectors with an outsized number of  non-zero tokens 
(8.9x longer document vectors compare to one of the SPLADE variance~\cite{2023SIGIR-Qiao}).
This is partly due to LLaMA-3's 128,256-token vocabulary, which is 4.2$\times$ larger than BERT's.
Although Lion-SP achieves state-of-the-art performance on BEIR and MS MARCO, it remains challenging to substantially improve its sparsity while preserving the relevance advantage provided by LLaMA-3.
Crucially, sparsity directly affects retrieval efficiency. In sparse retrieval, query processing latency and inverted-index storage generally scale with the number of non-zero terms in the query and document representations.

Previous research on static inverted index pruning~\cite{2001SIGIR-Carmel,2007SIGIR-Blanco, 2006CIKM-Buttcher,2023SIGIR-SPLADE-pruning} 
has studied term-centric and document-centric strategies to trim a term posting list or a document vector by a hard limit. 
Top token selection~\cite{2021-Yang-Masking,zhou2024retrievalbaseddisentangledrepresentationlearning} 
limits  the highest activated weight counts uniformly across all documents. 
Mass ratio pruning~\cite{2026arXivSINDI} is a contemporary effort to preserve top entries using the vector mass~\cite{2024SIGIR-SparseApproximate}. 
Hybrid thresholding (HT)~\cite{2023SIGIR-Qiao}
exploits a learnable term-centric thresholding architecture to filter out unimportant neural weights. These existing techniques offer limited learnability and insufficient fine-grained context-aware adaptivity, which limits their ability to produce substantially sparser LLM-driven representations.

To address this challenge, this paper proposes AdaSparse, short for "Adaptive Sparsification". This scheme achieves fine-grained pruning optimization through a synergy of adaptive strategies, including learnable soft per-vector top-$K$ and  per-term thresholding, which work in complement with FLOPs regularization.
A key distinction of this soft top-$K$ technique is that it avoids explicit, hard pruning after the $K$-th ranked term of a query or document
vector. This mechanism allows the model to capture sparsity penalty signals and autonomously adjust term weights during refinement.
AdaSparse is evaluated with Lion-SP for the MS MARCO passage dataset in training and testing,  and for BEIR datasets in zero-shot retrieval.
The experiments show that AdaSparse can generate much shorter sparse query and document vectors  by a 4x-5x reduction in
expanded query and document lengths while retaining highly competitive relevance.
It can reduce  actual retrieval latency  by 3.8x
and storage space by 4.2x using  a state-of-the-art  Seismic retriever~\cite{2024SIGIR-SparseApproximate}.

%% file: sigir26background.tex
\section{Background and Related Work}
\label{sect:background}

{\bf Learned sparse retrieval}.
Given a query, document retrieval finds top-$k$ ranked results by computing the similarity
between a query vector and a document vector. These query vectors are generated during online inference, while document vectors are pre-generated during offline index processing based on a learned neural transformer-based model. 

We use ``tokens'' and 	``terms'' interchangeably to describe the units produced by a language model's tokenizer.
The SPLADE family~\cite{Formal2021SPLADE, Formal2021SPLADEV2,Formal_etal_SIGIR2022_splade++} expands given query and document vectors 
to address vocabulary mismatch due to the lack of semantically-related terms in the original text.
SPLADE predicts term importance by mapping an input query or document sequence $S$ into a high-dimensional sparse vector 
$\mathbb{R}^{|V|}$, where $V$ is the size of the vocabulary. These models 
project the transformer-based contextualized representations multiplied by the embedding table 
of the transformer to obtain intermediate weight $w_{i,j}$, where $w_{i,j}$ denotes the importance of $i$-th input term in $S$ 
for each $j$-th term  $t_j$ in $V$. To produce the final sparse vector, log-smoothing and a ReLU activation are applied to ensure non-negativity, followed by a max-pooling operation across all input positions:
\begin{equation}
\label{eq:splade}
w_j =  \max_{i \in S}  \log(1 + \text{ReLU}(w_{i,j}))
\end{equation}
The final rank score  for each document during retrieval
is a dot product between query and document vector representations. For sparse vectors with many zeros, 
online retrieval can utilize 
a data structure called inverted index for fast low-cost inference~\cite{Mallia2017VBMW,Mallia2021deepimpact}.

The loss function of SPLADE models~\cite{ Formal2021SPLADE, Formal2021SPLADEV2,Formal_etal_SIGIR2022_splade++}
contains a ranking loss $L_R$ and FLOPs-based sparsity regularization~\cite{Paria2020Minimizing}.
The ranking loss has evolved from a log likelihood-based function  for maximizing positive document probability
to Margin-MSE~\cite{Hofsttter2020marginMSE}
 and/or KL Divergence for knowledge distillation.
Lion~\cite{zeng2025scalingsparsedenseretrieval} adapts the LLaMA architecture with bi-directional
attention following LLM2Vec~\cite{behnamghader2024llmvec} and trains dense or sparse retrieval models with knowledge distillation. 
This paper focuses on its sparse model called Lion-SP, which adopts
a formula similar to SPLADE with the FLOPs regularization in the loss function for representation sparsification.

The rank loss with FLOPs regularization used in Lion-SP  and SPLADE is summarized as follows.
Let $B$ be a set of training queries with $N$ documents involved in a batch. The overall loss for this batch including
$L_{R}$, the rank loss, with knowledge distillation is:

\begin{equation}
\label{loss}
L = (\frac{1}{|B|}\sum_{q \in B} L_R) +\lambda_Q L_Q + \lambda_DL_D 
\end{equation}
where $L_Q$ and $L_D$ are FLOPs for queries and documents:

\begin{equation}
 L_Q = \sum_{t_j \in V}  (\frac{1}{|B|}\sum_{q \in B}  w_j^q)^2; \ \ 
L_D =\sum_{t_j \in V}( \frac{1}{N} \sum_{ d =1}^{N} w_j^d  )^2. 
\end{equation}

{\bf Related work}. 
Other  sparsification studies ~\cite{2023SIGIR-Qiao,
2021-Yang-Masking,zhou2024retrievalbaseddisentangledrepresentationlearning, 2026arXivSINDI, 2024SIGIR-SparseApproximate,
lassance2022efficiency} will be discussed in next section. In addition to SPLADE, 
there are other learned sparse retrieval models 
~\cite{Zamani2018SNRM, Dai2020deepct,
Mallia2021deepimpact, Lin2021unicoil,2021NAACL-Gao-COIL,
shen2023lexmae} mainly based on BERT.
Sparse autoencoder studies interpret neural networks by decomposing hidden representations into 
disentangled sparse features~\cite{2024ICLR-SAE-Huben,bricken2023monosemanticity} with
a fixed sparsity constraint using top $K$ selection when optimizing representation
reconstruction~\cite{2025EMNLP-ASE-train,2025EMNLP-SAE-dense-Park,2024ICLR-SAE-Huben}.
Sparsification in VDR~\cite{zhou2024retrievalbaseddisentangledrepresentationlearning} for text-to-text and cross-model retrieval
uses a hard top-K limit while considering the inclusion of original words.
CSurF~\cite{2023ICTIR-CSurF-CMU} studies contextual sparse representation, which accounts for the existence of original words.

%% file: sigir26design.tex
\section{Design Considerations}
\label{sec:design}

Sparsification for Lion-SP faces a distinct challenge due to the massive LLaMA-3 128K vocabulary, which is 4.2$\times$ larger than that of BERT-based SPLADE. This larger token space makes it harder to distinguish and suppress unimportant terms. More importantly, the vocabulary composition of LLaMA-3 further increases representation redundancy.
First, the LLaMA-3 vocabulary assigns separate tokens to different variants of the same English word~\cite{reif2025vocabdietreshapingvocabulary}. Its tokenizer tends to produce more variants than BERT,  such as case variations. As a result, when a word is considered relevant, multiple variants of that word may also be marked as relevant. This makes the resulting query or document vector unnecessarily dense.
Second, LLaMA-3 is pre-trained for multilingual conversation, whose objective is not fully aligned with our retrieval setting. As a result, the multilingual knowledge acquired during pre-training may cause the model to activate semantically related non-English tokens even in an English-only setting, unnecessarily increasing the length of the generated sparse vectors.

The above characteristics of LLaMA-3 likely cause the sparse vectors generated by Lion-SP to contain many low-weight, semantically redundant term scores.  
As illustrated in Figure~\ref{fig:weight_dist}, a large fraction of the weights generated by Lion-SP on MS MARCO passages are small and similar in value. 
Table~\ref{tab:example} compares  a randomly sampled sparse query vector generated by SPLADE,
Lion-SP-1B, and  our work AdaSparse applied to Lion-SP-1B. Columns 3 and Column 4 are the number of variants generated by each model for keywords "years" and "governor" in this query.
SPLADE with BERT generates a vector of length 20, containing two variants of ``governor" (singular and plural).
Lion-SP-1B generates a vector of length 272 containing
9 English variants of the query word "year" (singular, plurals, uppercase, lowercase, having space before the term or not)
and two Chinese terms corresponding to  "year". Compare to SPLADE, it has a wider weight value range and a lower average value.
AdaSparse produces a condensed version with  4 English variants of "years" and 3 variants of "governor", while also shrinking the term weight
range and yielding an average term score closer to that of SPLADE.

\begin{table}[htbp]
    \centering
    \small
    \caption{Redundancy and weight range  of a sparse vector generated for query
``how many years did william bradford serve as governor of plymouth colony?''} 
    \begin{tabular}{l|c|c|c|c|c}
        \textbf{Model} & {\bf Length} &\textbf{\#years} &\textbf{\#governor} & \textbf{Min-Max} & \textbf{Avg}\\ 
\hline
SPLADE & 20 & 1  & 2   &  0.03 - 1.15  &0.40\\
\hline
Lion-SP-1B & 272 & 11 & 8 & 0.01 - 1.28 &0.20 \\
\hline
AdaSparse-1B &64 & 4& 3& 0.07 - 1.27& 0.41 \\
\hline
    \end{tabular}
    \label{tab:example}
    \end{table}

To bridge the gap between the strong relevance of sparse representation generated by Lion-SP and 
the efficiency constraints of online systems, we consider several representative sparsification strategies that can reduce representation size after training. Table~\ref{tab:pruning_comparison} compares the pros and cons of the four main previously proposed sparsification methods in its first four rows, which is elaborated below:
\begin{itemize}[leftmargin=*]
\item  {\bf Fixed top term selection}~\cite{2021-Yang-Masking,zhou2024retrievalbaseddisentangledrepresentationlearning} 
retains a constant number or percentage of terms for each input query or document vector.
While computationally efficient, these two top term selection methods do not account for context-aware weight distribution and their impact on relevance.
For example, it  may over-prune a complex, multi-faceted technical query (losing vital expansion terms) 
while under-prune a simple navigational query containing noisy and irrelevant terms.
The  $\alpha$-mass cutoff~\cite{2024SIGIR-SparseApproximate} is proposed to 
retain the smallest subset of top entries  whose cumulative sum reaches fraction $\alpha$  of the $L_1$ norm
of the summary vectors  of indexed clusters. 
This concept is extended recently to Mass Ratio Pruning (MRP)~\cite{2026arXivSINDI} for more adaptive top term  selection of document and query vectors.
This mass ratio  is  a fixed hyperparameter, and the pruning scope for each vector is not learnable. 

\item {\bf  FLOPs}~\cite{Paria2020Minimizing}, as used in Lion-SP~\cite{zeng2025scalingsparsedenseretrieval}
, regularizes based on average term weights in a training batch.
While effective for overall macro-level efficiency by shrinking large noisy weights,
it does not consider the relative semantic importance of terms at a fine-grained level.
Another limitation is that the FLOPs regularizer can produce very small gradients for terms with already small weights, making it difficult to further suppress these weights to zero. Thus, FLOPs tends to leave
a long tail of very small weights unless FLOPs regularization is applied heavily, which in turn affects relevance.
An  efficient version of SPLADE~\cite{lassance2022efficiency} studies several optimization tradeoffs including
$L_1$ regularization for query vectors while keeping FLOPs for documents.

\item  {\bf Learnable  uniform term thresholding (HT~\cite{2023SIGIR-Qiao})}
uses two learned thresholds for queries and for documents respectively,
and prunes a term when its weight is below such a global threshold.
This addresses the weakness of FLOPs in handling very small term weights. However, the distributions of neural weights vary substantially across terms, using a global threshold is too coarse-grained to handle such diversity effectively.
This limitation is further amplified in zero-shot retrieval, where applying a global threshold tuned on one domain to another domain may suppress all terms in some vectors, resulting in empty representations.

\end{itemize}

\begin{table*}[h]
    \centering
    \small
    \caption{Comparison of Query Pruning Methodologies}
    \label{tab:pruning_comparison}
    \begin{tabular}{@{}llll@{}}
        \toprule
        \textbf{Method} & \textbf{Mechanism} & \textbf{Strength } & \textbf{Weakness } \\ \midrule
        FLOPs   & Lower squared avg. term weight sum & Quickly shrink large noisy weights & Leave long tails of small weights\\

        Top-$K$ & Vector-centric, fixed $K$ or $K\%$  &Constant sparsity& Not learnable $K$. Over-prune complex vectors\\
        Mass ratio pruning & Vect.-centric. Top terms with mass ratio  & Mass-adaptive sparsity &  Not learnable mass ratio\\

        Uniform thresholding & Term-centric, $w_j > \tau$ & Relative importance filtering & Nonadapt. to distribut. diversity. Empty vect. risk \\
        Per-term thresholding &Term-centric,  $w_j > \tau_j$ & Individualized, value-adaptive& Difficult for flat distribut.\&reaching high sparsity\\ 
        Learnable soft top-$K$ &Vector-centric, penalty-driven soft  limit  & Adaptive without explicit pruning & Might be impacted by noisy term rankings \\
        \bottomrule
    \end{tabular}
\end{table*}

With the above in mind, we identify three design goals:
\begin{itemize} [leftmargin=*]
\item  {\bf Learnability for better adaptiveness}.
Sparsification needs to  incorporate  context-aware optimization that adapts to  distribution
variance in different query and document vectors for effective in-domain and out-of-domain  search. 
We seek a learnable top-$K$ method with a soft limit, tailored towards each query or document vector.

\item{\bf Fine-grained optimization}. Since many terms have small and similar values,
the proposed scheme also needs to address this problem in a pool of 128K distinct tokens. Given the weakness of uniform thresholding, 
we propose to extend the HT method by introducing individualized thresholds for each term, which enables more fine-grained
adaptation to term-specific weight distributions.

\item {\bf Multi-objective sparsification regularization}.
As discussed later and shown in the last two rows of 
Table~\ref{tab:pruning_comparison},  we find that the two proposed new methods 
 and the FLOPs method are complementary.
This suggests the need to design a multi-objective loss function through a synergy of multiple methods.

\end{itemize}

%% file: sigir26method.tex
\section{ The Proposed Methods for AdaSparse }
\label{sect:method}

Motivated by the design goals in Section~\ref{sec:design}, AdaSparse is built on two complementary sparsification techniques. To introduce adaptiveness into vector-level pruning, we propose STop, a learnable soft top-$K$ method that provides a context-dependent soft limit for each query or document vector. To improve fine-grained pruning of low-weight terms, we introduce PTT, a per-term thresholding scheme that suppresses low-weight terms using individualized thresholds. These two techniques are complementary to FLOPs regularization: STop mainly controls the expansion scale, while PTT refines the 
pruning of weak terms included during generation.
This section presents the two sparsification techniques and discusses how to integrate them with FLOPs in sparsity regularization.

\subsection{STop: Learnable soft top-$K$}
\label{sect:stop}
STop allows the model to adaptively select top-K important terms based on the context. Unlike
fixed top term selection~\cite{2021-Yang-Masking,zhou2024retrievalbaseddisentangledrepresentationlearning, 2026arXivSINDI, 2024SIGIR-SparseApproximate},
STop does not explicitly prune terms during training. Instead, it introduces a differentiable top-K selection signal that penalizes terms beyond a context-dependent soft limit, while allowing the model to adjust the effective number of retained terms for each vector. We now describe how this soft limit is defined, 
how the gating function is constructed, and used.

{\bf Step 1. Select a soft limit $K$}.
First, we set a soft limit for top-token selection for each vector: 
\[
	K = b \cdot |s|
\]
where $s$ is the input vector and $|s|$ denotes its length.
$b$ is a hyperparameter that determines the soft expansion factor.
This ensures that the expansion of  a document or a query is  reasonably rich to provide
a good bridge in query-document semantic matching. 
In practice, the actual expansion ratio may be smaller or slightly larger than the $b$ value used, as shown in Section~\ref{sect:eval}.

{\bf Step 2. Define a gating function for a document or query  vector}.
Using the soft limit defined above, we introduce a sigmoid-based gating function $G_j$ that compares each term weight with the weight around the K-th ranked position. This function produces a differentiable soft mask, which penalizes terms expanded beyond the soft top-K boundary during training:

\begin{equation}
\label{eq:stop-gate}
    G (w_j, v) = \sigma \left( \frac{ \widehat{w}_{K}- w_j  - \beta \cdot \mathbb{I}(t_j \in v_{\text{orig}})} {\gamma} \right)
\end{equation}
where
\begin{itemize}[leftmargin=*]
\item The  sigmoid function ranges between 0 and 1:   $\sigma(x) = \frac{1}{1+ e^{-x}}.$

    \item \textbf{$w_j$}: The raw importance score for term $t_j$ generated by the LLM expansion head.
    \item \textbf{$\widehat{w}_{K}$}: The {\bf rank-relative pivot} which 
is the average term weight at the $K$-th and $K+1$-th ranked positions. 
Namely $\widehat{w}_{K}= (w_K +w_{K+1})/2$. 
It acts as a floating limit that adapts to the vector's specific semantic density.
    \item \textbf{$\gamma$}: A temperature parameter that controls the steepness of the gating boundary.
    \item \textbf{$\beta$}: A positive scalar called the \textbf{lexical provenance bias} that ensures terms found in the 
original, unexpanded vector $v_{\text{orig}}$ have a significantly higher probability that the gating function returns  
a 0 or small value. 
    \item \textbf{$\mathbb{I}(t_j \in v_{\text{orig}})$}: An indicator function that returns 1 if term $t_j$ was part of 
the original vector and 0 otherwise.
\end{itemize}

\begin{figure}
  \includegraphics[width=0.8\columnwidth,keepaspectratio]{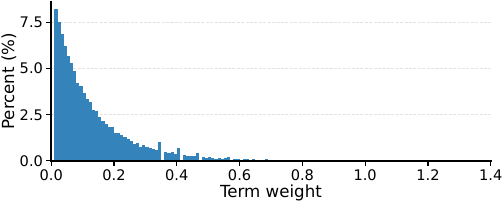}
   \caption{Term score distribution of Lion-SP 1B}
   \label{fig:weight_dist}
   \vspace{-2em}
\end{figure}

We choose hyperparameter $\beta$ to be bigger than the pivot value of all query and document vectors, namely $\beta >  max(\widehat{w}_K)$, 
so that original terms are effectively exempted from the gating penalty. Even at the beginning of the training,  $\widehat{w}_K$ 
may exceed $\beta$, $\widehat{w}_K$ value decreases gradually during iterative weight updating and the lexical bias can still make an impact.
Based on the term-weight distribution shown in figure \ref{fig:weight_dist}, where the nonzero scores of Lion-SP on MS MARCO are typically below 1.4, we set $\beta=2$.
Our training uses a scaling factor $\gamma=0.04$ to accelerate the steepness of the sigma function curve. 
Under these parameter choices, the gating function has the following properties.
\begin{itemize} [leftmargin=*]
\item We consider a term that $t_j$ appears in the top $K$ of the ranked term list of vector $v$;
or when a term $t_j$ appears in the original vector without expansion($\beta >  \widehat{w}_K$). 
The former condition gives $\widehat{w}_K < w_j$ while the latter gives $\widehat{w}_K < \beta$. Thus, the input of the gating function $
    \frac{ \widehat{w}_{K}- w_j  - \beta \cdot \mathbb{I}(t_j \in v_{\text{orig}})} {\gamma} <  0$ and the output value $G (w_j, v) \approx 0$. 
\item When $\widehat{w}_K \approx  w_j$ and  term $t_j$ is not in the original text before expansion,
then $G (w_j, v) \approx \sigma(0)  \approx 0.5$. This represents uncertainty of importance for terms 
without a clear decrease in value after the $K$-th position. 
\item Otherwise, 
$ \frac{ \widehat{w}_{K}- w_j  - \beta \cdot \mathbb{I}(t_j \in v_{\text{orig}})} {\gamma} >  0$, 
and the gating function approaches 1 rapidly with very small $\gamma$ factor.
$ G (w_j, v)  \approx  1$. 

\end{itemize}

{\bf Step 3. Apply  gating  function $G(w_j,v)$}.
 We only use it  for  model training in STop loss regularization $L_{top}$ defined as follows:
\begin{equation}
\label{eq:stoploss}
L_{top}=
\sum_{t_j \in V} \left( \frac{ \lambda_{qtop}}{|B|}  \sum_{q \in B} (w_j^q)^2 G(w_j,q)
+\frac{ \lambda_{dtop}}{N}  \sum_{d=1}^{N}  (w_j^d)^2  G(w_j,d) \right)
\end{equation}
where  $B$ is a set of training queries with $N$ documents involved in a batch.  
$\lambda_{qtop}$ and $\lambda_{dtop}$ are coefficient hyperparameters. 
The above formula  is essentially a gated $L_2$ loss, which will
be added to the overall regularization loss Formula~(\ref{eq:loss_final})  to be discussed shortly. 
       
The above design offers three distinct advantages:
\begin{itemize}[leftmargin=*]
\item \textbf{Loss-driven term weight refinement  without explicit top-$K$ pruning}. 
Gating function $G$ 
is not used to explicitly prune anything in training.
It sends a strong signal through  $L_{top}$ based on the three possible scenarios of the function value discussed above.  
Thus, the soft top-$K$ regularizer acts like a selective gravity. It is ``turned off" for the 
top-$K$ weights because $G(w_j,v) \approx 0$ for these weights. 
For other weights $w_j$ (soft position $K+1$ and below) with  $G(w_j,v) \approx 1$, 
which may trigger a decrease in  $w_j$ value during gradient-based weight updates.

\item {\bf Differentiability with value-based and count-adaptable soft limit}.
The above gating function is differentiable and the rank-relative pivot is value-based, which allows a soft limit with count-adaptivity, rather than a hard limit. 
As shown later in the gradient analysis of Section~\ref{sect:workflow}, 
that can lead to a negative gradient for some terms with $w_j \approx \widehat{w}_K$ 
below top $K$, contributing an opportunity for weight promotion.

\item \textbf{Lexical provenance}. 
This gating function naturally blends the idea 
of preserving   original tokens borrowed from  the 
previous  studies~\cite{2023ICTIR-CSurF-CMU,zhou2024retrievalbaseddisentangledrepresentationlearning}. 
Our configuration of $\beta$ yields  
$G(w_j,v) \approx 0$ when term $\mathbb{I}(t_j \in v_{\text{orig}})$,
leading to no loss penalty in $L_{top}$. 
Thus, the trained retrieval model captures this signal to keep the original terms, 
as original human-provided terms carry explicit intent.

\end{itemize}

{\bf Example}.
Given a  technical query: "quantum computing impact on finance", 
initial sorted weights are:  computing (0.5), portfolio (0.45), finance (0.45), impact (0.3), algorithm (0.3), 
risk (0.3), quantum (0.24), liquid (0.1).

For HT with a global query threshold of 0.26, any term with a weight below this value is pruned.
The final vector is \{ computing, portfolio, finance, impact, algorithm, risk \}.
This affects retrieval relevance, only matching documents containing these general terms. 
 The actual specific intent (Quantum) of the original query is lost.

For STop with $K=4$,  it will keep ``quantum'' because it appears in the original text.
It may also yield a negative gradient for ``algorithm'' and  	``risk'', and possibly boost them because they have the same 
weight as $\widehat{w}_K$=0.3 and $G(w_j,v)=0.5$. In fact, keywords ``algorithm'' and ``risk'' are related concepts in
option pricing and risk analysis algorithms that quantum computers can potentially speed up.
As a result, the model identifies core related keywords and   preserves intent  with  relevant expansions 
while still demoting ``liquid'' as a possible contextual noise or an irrelevant term.

{\bf Summary and limitation}. 
The learnable  soft top-$K$  can predict an optimized  number of terms  for each vector in 
a count-adaptive manner,
as stated in the last row of Table~\ref{tab:pruning_comparison}.
A weakness of this method is its sensitivity to noise in token importance rankings generated by the pre-trained and refined LLM. Under such distorted rankings, penalizing tokens beyond the soft position $K$ might negatively impact the preservation of useful words, causing brittle cliff behavior under noise.
Having a relatively large soft expansion factor $b$ minimizes the impact of this limitation.
STop also does not directly target the large number of low-weight terms produced by Lion-SP. This motivates the second component of AdaSparse, PTT, which performs fine-grained filtering of low-weight terms through individualized thresholds.

\subsection{PTT: Per-term thresholding}
\label{sect:PTT}
{\bf Pruning during document indexing and online query vector generation}. 
PTT prunes a term by setting its term weight $\widetilde{w}_j$ to zero, if its score falls below a certain threshold:
\begin{equation}
	\widetilde{w}_j = 
\begin{cases}
	w_j, \ \ \ \ w_j > \tau_{j} \\
	0,\ \ \ \ \ \  w_j \leq \tau_{j}
\end{cases}
\end{equation}
where $\tau_j$ is a term-specific threshold learned from the training process discussed below.
It can denote either $\tau_{D,j}$, the threshold used for document indexing,
or $\tau_{Q,j}$, the threshold used for online query vector generation.

{\bf Threshold learning during training}.
During training, as depicted in Figure~\ref{fig:workflow}(a), 
we use  the following thresholding function for  both queries and documents for smooth gradient propagation:
\begin{equation}
\label{eq:PTTthresholding}
	\widetilde{w}_j 
= w_j \cdot \sigma(\frac{w_j - \tau_j}{\pi})
\end{equation}
\begin{itemize}[leftmargin=*]
\item $w_j$: original weight for term $t_j \in V$ under vocabulary $V$.
\item $\pi$ controls the steepness of the sigma function curve in approximating pruning, 
allowing soft gradients during training and hard pruning during inference. We follow the setting of HT with $\pi$=0.04.
\item $\tau_{j}$: the threshold for term $t_j$. Query and document have two distinct sets of term threshold values.

\end{itemize}

The regularization loss leveraging the above individualized  thresholding for queries and documents is
defined as follows, generalized from the work of HT~\cite{2023SIGIR-Qiao}. 

\[
L_T = \frac{1}{|V|} \sum_{j \in V} [ \log(1 + e^{-\tau_{D,j}}) + \log(1 + e^{-\tau_{Q,j}})] .
\]
where $\tau_{D,j}$ and $\tau_{Q,j}$ are the thresholds learned for term $t_j$'s of documents and queries, respectively.

{\bf Summary and limitation.}
PTT  offers an individualized value-adaptive threshold for pruning to address the weakness of 
the HT method with finer granularity.  However, it still has an important limitation.
As stated in Row 5 of Table~\ref{tab:pruning_comparison},
if the weight distribution of documents for the same  term is extremely "flat," 
too many documents still keep such a term without effective pruning. 
Our evaluation finds that when we place a large regularization parameter $\lambda_{T}$
for using PTT  in the Lion-SP's loss function, 
the sparsity of vectors hardly increase after certain training steps.
Compared to PTT, the  learnable  soft top-$K$ method discussed in Section~\ref{sect:stop}
is more effective in
addressing the above sparsity resistance issue.

\subsection{Multi-objective sparsity regularization}
\label{sect:workflow}

We discuss how per-term thresholding and the learnable soft top-$K$  mechanism are integrated into the model training.
For a batch of training queries $B$, the original loss in Formula(\ref{loss}) 
is extended as: 
\begin{equation}
\mathcal{L} = (\frac{1}{|B|}\sum_{q \in B} L_R) +\lambda_Q L_Q + \lambda_DL_D + \lambda_T L_T
+ L_{top}. 
\label{eq:loss_final}
\end{equation}

$L_R$ is the rank loss following Lion-SP's setup. 
The sparsity regularization portion in  Formula(\ref{eq:loss_final}) has multiple
 learning objectives: FLOPs-based losses $L_Q$ and $L_D$ are the same as the ones in Formula (\ref{loss}) used in Lion-SP. 
$L_T$ is the regularization term corresponding to PTT, controlled by hyperparameter $\lambda_T$. 
Loss term $L_{top}$ is defined in Formula (\ref{eq:stoploss}) and introduced by STop. 
$\lambda_Q$, $\lambda_D$, $\lambda_T$, $\lambda_{qtop}$ and $\lambda_{dtop}$ 
are hyperparameters for component weighting.
\begin{figure}[ht]
    \centering
    \includegraphics[width=0.95\columnwidth]{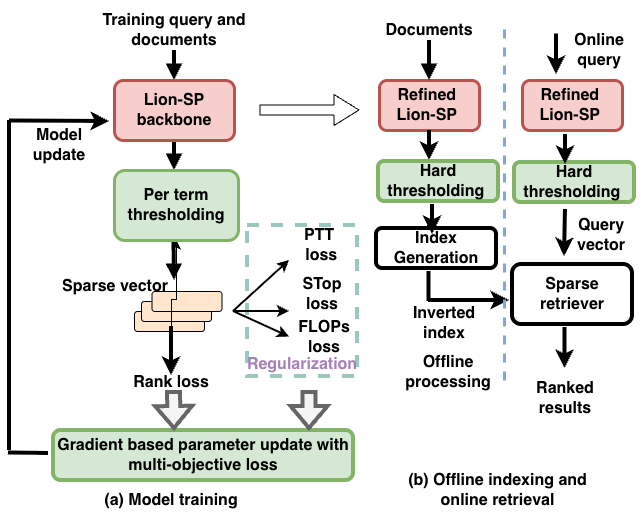}
    \caption{Offline and online processing with AdaSparse}
    \label{fig:workflow}
\end{figure}

{\bf Workflow}.
Figure~\ref{fig:workflow} illustrates the workflow of the AdaSparse framework. During training, AdaSparse adds two regularization components on top of Lion-SP training. The sparse representation generated by the Lion-SP backbone first goes through
the per-term thresholding (PTT) module that attenuates low-weight terms to near-zero using Formula (\ref{eq:PTTthresholding}).
We then compute the soft top-$K$ selection signal described in Section~\ref{sect:stop} and compute the overall loss, which consists of the rank loss, the FLOPs sparsification regularization terms, and two regularization components introduced in this work. The gradients of this loss are then computed, and the trainable model weights are updated accordingly.

During inference, AdaSparse drops terms whose weights fall below the threshold and then follows the same encoding procedure as Lion-SP, without applying any explicit top-$K$ pruning. The encoded sparse document vector will be indexed into an inverted index. The query vector, which is encoded in the same way, will be used to retrieve the relevant documents from the index.

\begin{table}[tbhp]
    \centering
    \small 
    \caption{Downward force gravitation on weight $w_j$ by each of the sparsification losses in AdaSparse at  a training step}
    \label{tab:gradient}
    \begin{tabularx}{\linewidth}{|X|l|X|} 
        \hline
        Targeted term weight $w_j$  & Gradient & Effect on $w_j$ \\
        \hline
        \multicolumn{3}{|l|}{\bf FLOPs with $L_{Q}$ and $L_D$} \\
        \hline
        All terms &  $2 \overline{w_j}  w_j$ & Downward force \\
        \hline
        \multicolumn{3}{|l|}{\bf Soft top-$K$ with $L_{top}$} \\
        \hline
        Soft top $K$ ($w_j\ge \widehat{w}_K$) & $\approx 0$ & Minimum impact \\
        Original term $t_j \in v_{orig}$ & $\approx 0$ & Minimum  impact \\
        Below top $K$, $w_j <\widehat{w}_K$ & $\approx 2 w_j $ & Downward force \\
        Below top $K$,  $w_j \approx \widehat{w}_K$ & $\approx  w_j  (1 -  \frac{ w_j }{4\gamma})$
   & Upward for  non-small  $w_j$ otherwise downward \\
        \hline
        \multicolumn{3}{|l|}{\bf Per-term thresholding with $L_{T}$} \\
        \hline
        $w_j \ge \tau_{Q,j}$ or $\tau_{D,j}$ & 0 & No update impact \\
        $w_j < \tau_{Q,j}$ or $\tau_{D_j}$ & 0 & 
$ \widetilde{w}_j \approx$ 0 by Formula (\ref{eq:PTTthresholding}) \\
        \hline
    \end{tabularx}
\end{table}

{\bf Gradients for downward weight updating during training}.
Given a term  weight $w_j$, 
Table~\ref{tab:gradient} summarizes downward force gradients for targeted term weights of vectors in
AdaSparse, yielded from each of the three sparsification techniques.

For FLOPs, let $L$ be  either $L_Q$ or $L_D$ and $\overline{w_j}$ is the average weight of term $t_j$ of
queries or documents within a training batch: 
\[
\frac{\partial L }{\partial w_j} = 2 \overline{w_j} w_j. 
\]

For soft top-$K$, let  $L$ be $L_{top}$ scaled down by $\lambda_{qtop}$ when $w_j$ is a query term
or by $\lambda_{dtop}$ when $w_j$  is a document term. 
\[
\frac{\partial L }{\partial w_j} = 2w_j G(w_j,v) +w_j^2 G(w_j,v)(1-G(w_j,v)) \frac{-1 }{\gamma}.
\]
There are three cases:
\begin{itemize}[leftmargin=*]
\item  For a term $t_j$ whose weight is above the pivot $\widehat{w}_K$, or for a term that belongs to the original vector before expansion, the gate satisfies $G(w_j, v) \approx 0$. Consequently, the corresponding gradient is approximately zero.
\item 
Otherwise  $t_j$ is not in top-$K$ and if  
$w_j < \widehat{w}_K$ ,$G(w_j, v) \approx 1$, and 
\[
\frac{\partial L }{\partial w_j} 
= w_j G(w_j, v) (2 -  \frac{ w_j (1-G(w_j,v))}{\gamma})
\approx 2w_jG(w_j, v) \approx 2w_j.
\]
\item 
If  $t_j$ is not in top-$K$ but $w_j \approx  \widehat{w}_K$, then $G(w_j, v) \approx 0.5$.
\[
\frac{\partial L }{\partial w_j} 
= w_j G(w_j, v) (2 -  \frac{ w_j (1-G(w_j,v))}{\gamma})
\approx w_j  (1 -  \frac{ w_j }{4\gamma}).
\]
When $w_j \leq 4 \gamma$, 
$\frac{\partial L }{\partial w_j} \ge 0$.
Otherwise $\frac{\partial L }{\partial w_j} < 0$. There is an opportunity for such a term weight below top $K$ to be increased
if the rank loss gradient supports. 
Notice our evaluation sets $\gamma=0.04$ and when  $w_j > 0.16$ and $w_j \approx  \widehat{w}_K$,
term $t_j$ has a second chance to be weighted more.

\end{itemize}
For PTT, the gradient with respect to query or document weight $w_j$ is always 0 while this weight
is re-scaled at each training step when the corresponding term threshold $\tau_{Q,j}$ or $\tau_{D,j}$ is updated.

{\bf Choices of hyperparameters for loss weighting.}
We consider $L_T$ equally important as rank loss, and thus $\lambda_T=1$. 
We let $\lambda_{qtop}=\lambda_Q$ and  $\lambda_{dtop}=\lambda_D$
so that the gradients from FLOPs and STop are on a comparable scale.
This allows the nonzero gradient contribution from STop  to 
become significant in updating  weight $w_j$.

{\bf Synergy of the three techniques and small term weight handling}. 
The tripartite sparsity regulation structure in AdaSparse outperforms standalone methods or bipartite combinations 
in terms of relevance and/or sparsity through a self-correcting feedback loop discussed as follows. 
The rank loss suppresses weights that hurt ranking accuracy  while preserving or promoting weights that improve it. 
FLOPs applies a global downward  pressure to term weights and is effective at shrinking large noisy activations, but its 
gradient can become weak for already small weights. Complementary to FLOPs, STop provides a vector-adaptive 
sparsity signal. As shown in the above gradient analysis,  if 
term weight $w_j$ exceeds the vector's pivot value or $w_j \in v_{orig}$,
$G(w_j, v)\approx0$, STop contributes little additional  penalty and the term is mainly governed by the rank loss and FLOPs. 
Once a term falls below the pivot value, even if {\bf such weight can be small}, 
its gradient is amplified by the STop. This accelerates the decrease of the weight toward zero. If such a term is important for relevance, the rank loss 
can counteract this sparsity pressure and promote it back among the soft top-$K$ terms. PTT identifies small term weights and approximates their removal by scaling them down with a sigmoid function, where the scaling is controlled by updated term-specific thresholds.

When the sparsity regularization  only uses one or two of the three methods, there are weaknesses as listed in the 
4th column of Table~\ref{tab:pruning_comparison}. AdaSparse addresses them  in a collective manner. 
Table~\ref{tab:problemcases} summarizes the names of the main techniques activated for each sparsification scenario, and some problem cases have been
discussed in Section~\ref{sec:design}. 
For {\bf  small term weight $w_j$,}
PTT is effective to prune in general but is less effective when there is  {\bf a relatively
flat distribution} across documents. 
In that case, FLOPs is also less effective when  $0 <w_j<1$ because  its gradient 
$2 \overline{w_j}  w_j$ can be small with $0< \overline{w_j}  <1 $. 
Since $\lambda_{qtop}=\lambda_Q$ and $\lambda_{dtop}=\lambda_D$, 
STop can contribute the most significant gradient value 
with its gradient  $2 w_j$ when $G(w_j,v)$ is close to  1. 

\begin{table}[tbhp]
        \centering
        \caption{
Main techniques applicable to term weight scenarios}
\label{tab:problemcases}
        \begin{tabular}{|l|l|}
        \hline
        Scenarios  & Main techniques \\
        \hline
        \hline
	Relevant and irrelevant  weights & Rank loss  \\
        \hline
 	Large unimportant weights & FLOPs \\
 	Noisy term weight rankings  & \\
        \hline
 	Small unimportant weights & Per-term threshold. \\
 	Diverse term weight distribution  & \\
        \hline
 	Small weight with flat distribution  & Soft top $K$  \\
 	Excessively long expansion&  \\
        \hline
        \end{tabular}
\end{table}

%% file: sigir26experiment.tex
\section{Evaluation Studies}
\label{sect:eval}

\subsection{Settings}

\noindent
{\bf Datasets and relevance metrics.}
The MS MARCO dataset~\cite{bajaj2018msmarcohumangenerated,Craswell2020OverviewOT}  with about 8.8M passages is used 
for model training. Measured using Huggingface LLaMA-3.2 tokenizer and excluding the beginning special token, the average length of the original queries before expansion in the MS MARCO passage dataset is approximately 8 words, while the average passage length is 76. 
The performance    evaluation for in-domain search is conducted  on its passage Dev set with 6,980 queries, 
TREC Deep Learning (DL) 2019 set with 43 judged queries, and DL 2020  sets with 53 judged queries.
Following the common practice, we report the mean reciprocal rank (MRR@10) for Dev, 
and report NDCG@10 for TREC DL sets.

We also evaluate 
zero-shot retrieval
performance on the BEIR collection which contains 13 datasets~\cite{thakur2021beir} with
an average length of original queries
ranging between 5 and 245 words, and the average original document length ranging between 14 and 369 words.  

\noindent
{\bf Model training and configurations.}
By default, we use the LLaMA-3.2 with 1 billion parameters as the primary LLM and 
we also report the performance with 8 billion parameters.
We build on the Lion-SP codebase~\cite{lionRepo} and will release our implementation on GitHub\footnote{\url{https://github.com/ViViVidam/AdaSparse}}.
The model is trained with a combination of contrastive loss (CL) and knowledge distillation (KD), following Lion-SP, and has two steps:
1) We pre-train the LLaMA-3.2 model on the MS MARCO corpus using a masked next-token prediction objective.
2) We then fine-tune the model on MS MARCO passages using teacher-generated scores for KD and hard labels for CL.

Pretraining is  conducted  on 2 NVIDIA H100 GPUs, with a per device batch size of 32. 
Finetuning on MS MARCO is conducted on  4 NVIDIA H100 GPUs with 1 epoch.
When comparing different loss functions for the Lion-SP refinement, we always start from the same pretrained model, 
and only change the loss functions and other necessary settings for a fair comparison across different methods.

\noindent
{\bf Sparsity and latency metrics.} 
We report the average lengths of the expanded query and document vectors generated by the AdaSparse and the baselines.
Shorter lengths typically imply shorter retrieval latency on average,
and thus, query and document lengths serve as a proxy for retrieval time cost. 
We will study this relationship in more detail and report the actual latency 
of learned representations 
when the  corresponding inverted indices are executed by a state-of-the-art Seismic 
retriever~\cite{2024SIGIR-SparseApproximate} at  retrieval depth 10 and 1,000. Because Seismic only supports unsafe search, the comparison is performed under approximately matched effectiveness loss relative to safe retrieval.
This evaluation is conducted on AMD EPYC 9R45 Processor CPU with 256GB of memory. 
The latency reported is the average of three runs to minimize variability.
\begin{table}[thbp]
	\centering
	\caption{Default AdaSparse hyperparameters used}
	\label{tab:hyperparameters}
		\begin{tabular}{l|l|l}
		\hline
Hyperparameters & Derivation  & Value setting\\
		\hline
\multicolumn{3}{l}{\textbf{Weights for learning loss components}}\\
		\hline
$\lambda_Q$ and   $\lambda_D$ for FLOPs  & Follow~\cite{zeng2025scalingsparsedenseretrieval}  & $\lambda_Q=0.05$, $\lambda_D=0.04$ \\
		\hline
$\lambda_T$ for PTT  & Follow~\cite{2023SIGIR-Qiao} 
 & $\lambda_T=1$ \\
		\hline
$\lambda_{qtop}$ for STop & 
$\lambda_{qtop}=\lambda_Q$  &  $\lambda_{qtop}=0.05 $ \\  
		\hline
$\lambda_{dtop}$ for STop & 
$\lambda_{dtop}=\lambda_D$ &  $\lambda_{dtop}=0.04$ \\  
		\hline
\multicolumn{3}{l}{\textbf{Parameters for sigmoid gating functions }}\\
		\hline
Soft expansion limit & Parameter & $b=10$ for query\\
& sweeping&  $b=8$ for doc.\\
		\hline
Lexical provenance& $\beta =  max(\widehat{w}_K) $  &  $\beta=2$\\
		\hline
Temperature for PTT & Follow~\cite{2023SIGIR-Qiao} 
& $\pi = 0.04$  \\
		\hline
Temperature for STop & $\gamma=\pi$ &  $\gamma = 0.04$  \\
		\hline
		\end{tabular}
\end{table}

\noindent
\textbf{Hyperparameter setting.}   
Table~\ref{tab:hyperparameters} lists hyperparameter values used in training the Lion-SP 1B and 8B models with AdaSparse. 
Column 2 lists how we derive each value.
We follow Lion-SP's setup for  $\lambda_Q$ and  $\lambda_D$
and HT's setup for $\lambda_T$ and $\pi$.  
For $\beta$, $\lambda_{qtop}$, and $\lambda_{dtop}$, our design constrains them 
as discussed in Sections~\ref{sect:stop} and ~\ref{sect:workflow}. 
We set $\pi=\gamma$ as both are  for temperature-like smoothing in sigmoid gating. 
For soft expansion factor $b$,  we sweep the hyperparameter value on the MS MARCO passage training data.  
Section~\ref{sect:ablation} gives a sensitivity study  when varying $b$ and  $\lambda_T$.

\vspace{-0.5em}
\subsection{A comparison with baselines}
\noindent
{\bf A comparison on relevance and sparsity on MS MARCO Dev and BEIR}.
Table \ref{tab:overall}  compares the performance of
trained Lion-SP using AdaSparse and 
other baselines.   
For the baselines, Lion-SP-1B and Lion-SP-8B use  official Lion-SP releases with 1B and 8B checkpoints.  
Lion-SP-1B-repro is our reproduction of model Lion-SP-1B trained with 1 epoch. 
Lion-SP-8B also uses 1 epoch in training.
FLOPs-1B  is one variation of Lion-SP that we produce with
 $\lambda_{Q}=0.6$ and $\lambda_{D}=0.4$ for queries and documents, respectively.
We have tried a number of variations of FLOPs-1B with 
different loss coefficient parameters for FLOPs loss to force the model to produce
 the average 
vector length close
to what AdaSparse-1B delivers. 
L1-FLOPs-1B uses   $L_1$ regularization for each query  and keeps FLOPs for documents in training Lion-SP, 
following~\cite{lassance2022efficiency}.

Top-$305$-1B uses $K=305$ following~\cite{2021-Yang-Masking}.
VDR-256-1B keeps  top 256 including original words following its setting~\cite{zhou2024retrievalbaseddisentangledrepresentationlearning}. 
MRP-1B listed uses $\alpha=0.75$ for both documents and queries,
which produces a sparsity level close to what  the AdaSparse setting delivers. We apply MRP as a post-processing 
method to the reproduced Lion-SP-1B model.
HT-1B uses the recommended HT's configuration~\cite{2023SIGIR-Qiao}. 

Table~\ref{tab:overall} shows that AdaSparse produces much shorter sparse vectors  while its relevance remains highly competitive to the original Lion-SP model. Compared to Lion-SP-1B-repro,  AdaSparse-1B shows negligible performance degradation ($\approx$0\% on MS MARCO Dev and 0.74\% on BEIR), 
while reducing query and document lengths to 31\% and 26.6\% of the original, respectively. Similarly, AdaSparse-8B exhibits less than a 0.3\% performance drop on MS MARCO Dev and $\approx$1.6\% degradation on BEIR, with an 82\% reduction in query length and a 79.6\% reduction in document length.

In general, AdaSparse outperforms other baselines in balancing sparsity and  relevance. 
Among all static pruning methods  with top term selection which can easily achieve a target sparsity level, 
MRP is more  competitive than Top-K and VDR in relevance.
Top-K and VDR with uniform pruning  for all vectors can remove meaningful terms from complex documents, hurting relevance. 
Even MRP's pruning is vector-specific based on the vector mass, its relevance is  still weaker  than  AdaSparse, especially for BEIR
with about 2.3\% lower  average NDCG@10  due to its non-learnable pruning. 
Compared to Lion-SP-1B-repro, 
HT-1B has a 1\% MRR degradation on MS MARCO Dev and 1.6\% on BEIR. 
Furthermore, its sparsity is suboptimal: the query length drops by only 9.1\%, and its document length is 39.6\% longer than that of AdaSparse-1B.

This table also shows that  Lion-SP/AdaSparse with 8B parameters compared to 1B can boost 
2.2\% for MS MARCO Dev, 1\% for DL19, 3.2\% for DL20, and 0.4\% for BEIR. 
Document vector length changes by only 6\%, while query length increases by 30\%, from 65 to 85.

\begin{table}[thbp]
	\centering
	\caption{A comparison of AdaSparse and baselines in relevance and sparsity (average vector length)  with LLaMA-3.2}
	\label{tab:overall}
	
	\resizebox{\linewidth}{!}{%
		\begin{tabular}{c|c|c|c|c|c|c}
		\hline
		& MS M. & DL19 & DL20 & BEIR & Query & Doc \\
		& MRR  & NDCG & NDCG & NDCG & Len & Len \\
		\hline
		Lion-SP-1B~\cite{zeng2025scalingsparsedenseretrieval} & {\bf 0.410} & 0.747 & {\bf 0.753} & 0.535 & 293 & 1250 \\
		Lion-SP-1B-repro & 0.408 & 0.753 & 0.744 & {\bf 0.545}  & 208 & 1052 \\ 
		FLOPs-1B & 0.400 & 0.745 & 0.743 & 0.530 & 130 & 315 \\
		L1-FLOPs-1B & 0.404 & 0.754 & 0.718 & 0.531 & 98 & 297 \\
		Top-305-1B & 0.372 & 0.709 & 0.690 & 0.231 & 85 & 308 \\
		VDR-256-1B& 0.398 & 0.741 & 0.717 & 0.493 & 103 & 272 \\
		MRP-1B (0.75)& 0.406 & 0.748 & 0.742 & 0.529 & 78 & \textbf{268} \\
		HT-1B& 0.404 & 0.754 & 0.744 & 0.536 & 189 & 391 \\
		\hline
		\textbf{AdaSparse-1B} & {0.408} & \textbf{0.755} & {0.729} & {0.541} & \textbf{65} & {280} \\
		\hline
		\hline
		Lion-SP-8B~\cite{zeng2025scalingsparsedenseretrieval} & {\bf 0.418} & 0.760 & 0.766 & {\bf 0.552} & 476 & 1457 \\
		\hline
		\textbf{AdaSparse-8B} & 0.417 & \textbf{0.763} & {0.752} & {0.543} & \textbf{85} & \textbf{297} \\
		\hline
		\end{tabular}%
	}
\end{table}

\begin{table}[thbp]
    \centering
    \caption{Latency and storage comparison of trained representations. Latency is measured in milliseconds (ms).}
    \label{tab:ada_latency}
    \renewcommand{\arraystretch}{1.2}
    \resizebox{0.95\columnwidth}{!}{%
        \begin{tabular}{lrrcccr}
            \toprule 
            \multirow{2.5}{*}{\textbf{Model}} & \multicolumn{2}{c}{\textbf{Latency (ms)}} & \multicolumn{3}{c}{\textbf{MRR@10 (Dev)}} & \textbf{Storage} \\
            \cmidrule(lr){2-3} \cmidrule(lr){4-6}
             & $k=10$ & $k=1000$ & $k=10$ & $k=1000$ & Safe & (GB) \\
            \midrule
            
            Lion-SP-1B      & 1.179 (4x)   & 9.048 (3.8x) & 0.407 & 0.408 & 0.410 & 138 (4.2x) \\
            Lion-SP-8B      & 1.341 (4.6x)  & 9.812 (4.1x)   & 0.415 & 0.416 & 0.418 & 154  (4.7x)  \\
            HT-1B   & 0.613 (2.1x)     & 3.223 (1.4x) & 0.402 & 0.403 & 0.404 & 58 (1.8x)  \\
            \midrule
            AdaSparse-1B    & 0.292 (1x)   & 2.390 (1x) & 0.401 & 0.406 & 0.408 & 33 (1x)  \\
            AdaSparse-8B & 0.317 (1.1x) & 2.863 (1.2x) & 0.413  & 0.415  & 0.417  & 34 (1x)  \\
            \bottomrule
        \end{tabular}%
    }   
\end{table}

{\bf Latency and storage cost}.
Table~\ref{tab:ada_latency} shows the latency with retrieval depth 1,000
and storage cost  of Lion-SP, AdaSparse and HT.
We use a uniform Seismic setting across all the datasets for the same model; however, 
different models might have different Seismic settings due to the length difference.
To run Seismic efficiently, a grid search on its parameter choice for MS MARCO is needed~\cite{2024SIGIR-SparseApproximate}. 
Our Seismic setting is chosen so that the MRR@10 on the MS MARCO passage Dev set is within 0.002 of safe search.
This is intended to align the approximation-induced effectiveness loss across methods, 
so that the latency comparison reflects end-to-end retrieval efficiency under comparable retrieval quality degradation. 
For AdaSparse, AdaSparse-8B and HT-1B models, we set Seismic's parameters $\alpha=0.4$, $\lambda=4000$,
given a similar document sparsity level. We use query cut 8 for HT-1B model, while 6 for
AdaSparse and AdaSparse-8B because HT-1B has a longer query length. 
Column 6  is MRR@10 of safe retrieval, and it is slightly higher than Columns 4 and 5 due to the retrieval approximation of Seismic.
The result shows there is a strong correlation between higher sparsity of a trained representation and shorter latency.
AdaSparse yields shorter document lengths on average and thus its inverted
index space is proportionally smaller while the online latency is much shorter
than HT-1B  and original Lion-SP.

%% file: sigir26ablation.tex
\begin{table*}[htbp]
\centering
\caption{Average query length (Q Len) and
document length (D Len) on MS MARCO Dev and BEIR datasets. $b$=10 for queries and 8 for docs. \textbf{Average row excludes MS MARCO Dev. Ratio in the average is the average number of all the ratios across datasets.}}
\renewcommand{\arraystretch}{1.2}
\resizebox{\textwidth}{!}{%
\begin{tabular}{lccrrcrrcrrcc}
\hline
 & \multicolumn{2}{c}{\textbf{Original without expansion}} & \multicolumn{3}{c}{\textbf{LION-SP-1B (FLOPs)}} & \multicolumn{3}{c}{\textbf{HT-1B (Hybrid Thresholding)}} & \multicolumn{3}{c}{\textbf{AdaSparse (FLOPs+PTT+STop)}} \\
\cline{2-3} \cline{4-6} \cline{7-9} \cline{10-12}
Dataset & Q Len & D Len & Q Len & D Len & relevance & Q Len  & D Len & relevance & Q Len & D Len & relevance \\
\hline
MS MARCO Dev       & 8   & 76    & 293 (36.6x)  & 1251 (16.5x) & 0.410 & 189 (23.6x) & 391 (5.1x) & 0.404 & 65 (8.1x)   & 280 (3.7x)  & 0.408 \\
\hline
TREC-COVID     & 15  & 235   & 349 (23.3x)  & 2212 (9.4x)  & 0.785 & 179 (11.9x) & 765 (3.3x)   & 0.829 & 85 (5.7x)   & 555 (2.4x)  & 0.817 \\
NFCorpus       & 5   & 342   & 290 (58.0x)  & 2586 (7.6x)  & 0.364 & 190 (38.0x) & 910 (2.7x)   & 0.370 & 63 (12.6x)  & 702 (2.1x)  & 0.371 \\
NQ             & 10  & 108   & 298 (29.8x)  & 1524 (14.1x) & 0.586 & 186 (18.6x) & 543 (5.0x)   & 0.581 & 71 (7.1x)   & 373 (3.5x)  & 0.579 \\
HotpotQA       & 23  & 70    & 502 (21.8x)  & 1347 (19.2x) & 0.692 & 222 (9.7x) & 505 (7.2x)   & 0.692 & 123 (5.3x)  & 311 (4.4x)  & 0.698 \\
FiQA-2018      & 13  & 169   & 398 (30.6x)  & 1740 (10.3x) & 0.378 & 169 (13.0x) & 552 (3.3x)   & 0.378 & 89 (6.8x)   & 433 (2.6x)  & 0.381 \\
ArguAna        & 245 & 208   & 2363 (9.6x)  & 2220 (10.7x) & 0.488 & 450 (1.8x) & 758 (3.6x)   & 0.487 & 324 (1.3x)  & 565 (2.7x)  & 0.470 \\
Touche-2020    & 8   & 369   & 337 (42.1x)  & 2005 (5.4x)  & 0.329 & 153 (19.1x) & 671 (1.8x)   & 0.318 & 70 (8.8x)   & 509 (1.4x)  & 0.348 \\
DBPedia-entity & 7   & 76    & 288 (41.1x)  & 1385 (18.2x) & 0.464 & 174 (24.9x) & 525 (6.9x)   & 0.446 & 69 (9.9x)   & 326 (4.3x)  & 0.452 \\
SCIDOCS        & 13  & 225   & 597 (45.9x)  & 2282 (10.1x) & 0.170 & 268 (20.6x) & 872 (3.9x)   & 0.169 & 124 (9.5x)  & 582 (2.6x)  & 0.170 \\
FEVER          & 11  & 125   & 672 (61.1x)  & 1554 (12.4x) & 0.870 & 237 (21.5x) & 607 (4.9x)   & 0.856 & 148 (13.5x) & 386 (3.1x)  & 0.878 \\
Climate-FEVER  & 25  & 125   & 958 (38.3x)  & 1554 (12.4x) & 0.304 & 285 (11.4x) & 607 (4.9x)   & 0.249 & 212 (8.5x)  & 386 (3.1x)  & 0.281 \\
SciFact        & 19  & 320   & 1021 (53.7x) & 2741 (8.6x)  & 0.734 & 321 (16.9x) & 997 (3.1x)   & 0.731 & 222 (11.7x) & 720 (2.3x)  & 0.727 \\
Quora          & 12  & 14    & 365 (30.4x)  & 393 (28.1x)  & 0.791 & 165 (13.8x) & 154 (11.0x)   & 0.866 & 82 (6.8x)   & 85 (6.1x)   & 0.858 \\
\hline
Average        & 31  & 184   & 649 (37.4x)  & 1811 (12.8x) & 0.535 & 231 (17.0x) & 651 (4.7x)   & 0.536 & 129 (8.3x)  & 456 (3.1x)  & 0.541 \\
\hline
\end{tabular}%
}
\label{tab:adaptive}
\end{table*}

\begin{table}[h]
\centering
\caption{Impact  of  hyperparameter changes in AdaSparse-1B}
\label{tab:combined_responsiveness}
\resizebox{\columnwidth}{!}{%
\begin{tabular}{lcccc|lcccc}
\toprule
\multicolumn{5}{c|}{\textbf{Different $b$ for query}} & \multicolumn{5}{c}{\textbf{Different $\lambda_T$}} \\
\cmidrule(r){1-5} \cmidrule(l){6-10}
\textbf{Setting} & \textbf{MS M.} & \textbf{Q Len} & \textbf{D Len} & \textbf{BEIR} & \textbf{Setting} & \textbf{MS M.} & \textbf{Q Len} & \textbf{D Len} & \textbf{BEIR} \\
\midrule
$b=10$ & 0.408 & 65 & 280 & 0.541 & $\lambda_T=1$ & 0.408 & 65 & 280 & 0.541 \\
$b=8$ & 0.408 & 59 & 273 & 0.539 & $\lambda_T=3$ & 0.408 & 58 & 257 & 0.539 \\
$b=5$ & 0.407 & 48 & 262 & 0.536 & $\lambda_T=6$ & 0.408 & 54 & 245 & 0.534 \\
\bottomrule
\end{tabular}
}
\end{table}
 
\noindent
{\bf Detailed Results on MS MARCO and BEIR datasets.}
Table~\ref{tab:adaptive} shows the average query and document lengths for each of 13 BEIR datasets and MS MARCO before and after expansion
when using AdaSparse and the original Lion-SP regularization loss (FLOPs).
The three columns marked as "Relevance"  list
 MRR@10 for MS MARCO Dev and  NDCG@10 for BEIR. The average number listed in the last row
is calculated on 13 BEIR datasets, excluding MS MARCO. Latency is in milliseconds. Soft scaling 
limit during model training is   $b=10$ for queries and 8 for documents. 
Token counting for lengths is based on the LLaMA-3 tokenizer for all columns. 
The actual expansion factor listed is the vector length after model expansion divided by the original length
for queries and documents, respectively.
We include results at the dataset level on BEIR to show that the model behavior is consistent in both out-of-domain and in-domain scenarios. 
The results in Table~\ref{tab:adaptive} suggest the following key observations:
\begin{itemize}[leftmargin=*]
\item {\bf AdaSparse vs. Lion-SP with FLOPs only regularization}.
AdaSparse shows strong robustness, maintaining high sparsity for expanded document and query vectors in both in-domain and out-of-domain settings. On the BEIR benchmark, AdaSparse reduces the average vector length to 20\% of Lion-SP's query length and 25\% of its document length.

\item {\bf AdaSparse-1B vs. HT-1B}.
AdaSparse's MRR@10 for MS MARCO Dev is 1\% higher than that of HT-1B, while query length is 2.9x shorter and the document length is 1.4x shorter.
Their relevance is comparable on DL'19 and DL'20.  
AdaSparse's BEIR NDCG@10 is 0.9\% higher than that of HT-1B  meanwhile
the overall sparsity yielded by AdaSparse is much higher than HT-1B.
These results demonstrate that our design effectively addresses the weakness of HT-1B  in filtering out low-weight terms
and managing the score distribution shift.
This allows AdaSparse to preserve relevance while achieving a more aggressive overall sparsity.

\item {\bf AdaSparse's expansion adaptiveness of query and document lengths on different datasets}.
The actual document expansion factors of all datasets are much smaller than the soft limit ($b=8$). 
For actual query expansion factors, 11 out of 14 datasets fall into the soft top-$K$ limit of AdaSparse with 
MS MARCO training ($b=10$).
The other 3 BEIR datasets have generated query lengths 11.7x to 13.5x longer than the original one. 
This also shows that AdaSparse is adaptive to the actual expansion need.
When semantically rich terms are needed for relevance, AdaSparse still expands queries beyond the trained soft limit $b$.

The number of non-zero entries generated by AdaSparse scales along with the input length with dynamic adjustments of the final sparsity level adaptive to the input length. AdaSparse generates the longest length 
on the ArguAna dataset, as ArguAna input is the longest among all. Meanwhile, AdaSparse generates the shortest vector on NFCorpus, 
as the input length is the shortest. 
\end{itemize}

\subsection{Sensitivity of hyperparameters $b$ and $\lambda_T$} 
\label{sect:ablation}
The left part of Table~\ref{tab:combined_responsiveness} shows the impact of varying soft expansion limit factor $b$ for queries from 10 to 5
while keeping soft limit  for documents as 8: the query length decreases accordingly, 
accompanied by a small drop in BEIR NDCG@10 score.  
Since the document and query encoders are  shared,  only varying the query limit $b$ also impacts document sparsity in some degree.
The right part of Table~\ref{tab:combined_responsiveness} shows the impact of increasing $\lambda_T$ from 1 to  6.
Increasing $\lambda_T$ also shortens query and document vectors and similarily 
overly aggressive sparsification affects relevance. 

Comparing the left and right parts of Table~\ref{tab:combined_responsiveness}, 
decreasing query limit  $b$ can yield sparser query vectors than  increasing $\lambda_T$  while  having a smaller  zero-shot performance 
drop and  a similar MS MARCO relevance. 
As marked  in Row 1 of Table~\ref{tab:adaptive} for MS MARCO,
AdaSparse expands 8.1x on average for queries under soft limit $10$ and expands 3.7x for documents under  limit $8$. 
Even though there is a gap between the actual expansion factor and targeted soft limit $b$,
varying $b$ still gives a much more direct and explicit  control of generated vector lengths than tuning other hyperparameters including $\lambda_T$.
Thus $b$ is chosen as our primary governing parameter to constrain the targeted sparsity while keeping $\lambda_T=1$ as shown in 
Table~\ref{tab:hyperparameters}.

\begin{table}[t]
\centering
\caption{Comparison under two sparsity configurations with 1B. Q/D Len denotes average query and document lengths. 
Relevance columns report MS MARCO Dev MRR@10 and BEIR average NDCG@10, respectively.}
\label{tab:TwoLengthConfig}
\scriptsize
\setlength{\tabcolsep}{3pt}
\resizebox{\columnwidth}{!}{
\begin{tabular}{lcccc}
\toprule
\multirow{2}{*}{Method} 
& \multicolumn{2}{c}{Smaller Config.} 
& \multicolumn{2}{c}{Default Config.} \\
\cmidrule(lr){2-3} \cmidrule(lr){4-5}
& Q/D Len & MS M./BEIR 
& Q/D Len & MS M./BEIR \\
\midrule
AdaSparse  
& 50 / 206 
& 0.409 / 0.537 
& 65 / 280 
& 0.408 / 0.541 \\

FLOPs      
& 100 / 187 
& 0.382 / 0.508 
& 130 / 315 
& 0.400 / 0.530 \\

L1-FLOPs 
& 104 / 221 
& 0.397 / 0.512 
& 98 / 297 
& 0.404 / 0.531 \\

MRP        
& 60 / 204 
& 0.394 / 0.526 
& 78 / 268 
& 0.406 / 0.529 \\
\bottomrule
\end{tabular}
}
\end{table}

{\bf Two sparsity configurations}. 
Table~\ref{tab:TwoLengthConfig} 
compares the performance of AdaSparse and few baselines  in two sparsity configurations. 
The default configuration of  AdaSparse uses hyperparameters from Table~\ref{tab:hyperparameters} while
the smaller  configuration is the same except that  the soft expansion limit factor is changed as  $b=6$ for queries and $b=5$ for  documents.
To match the above two configurations in training Lion-SP with 1B,  
we adjust the hyperparameters of other baselines for a similar average document length.
MRP uses  $\alpha=0.75$ for the default while $0.65$ for the smaller configuration.
FLOPs uses $(\lambda_q,\lambda_d)$
= (0.6,0.4) for  the default and (5,4) for the smaller.
L1-FLOPs uses $(\lambda_q,\lambda_d)$ = (0.005,0.5) for default
and (0.006,1.5) for the smaller.
The results show that 
the smaller configuration of AdaSparse retains  MS MARCO relevance of the default configuration with an almost identical MRR number
while having  a modest relevance drop for zero-shot retrieval with BEIR due to more aggressive sparsification. 
On the other hand, the smaller AdaSparse configuration still outperforms 
the relevance of other baselines in both smaller and default configurations at a similar or better sparsity level. 
\comments{
AdaSparse remains highly resilient on in-domain retrieval 
under a more demanding sparsity level, achieving the same performance on 
MS MARCO Dev. However, this more aggressive sparsification leads to a moderate 
drop on BEIR. Overall, the smaller AdaSparse configuration can still outperform both configurations of
other baselines in relevance at a similar or better sparsity level. 
}

\begin{table}[thbp]
    \centering
    \caption{Ablation studies with different combinations}
    \label{tab:effect_topk}
    \small
    \resizebox{\columnwidth}{!}{
    \begin{tabular}{l|ccc|c}
        \hline
        \multirow{2}{*}{Performance with 1B} &
        \multicolumn{3}{c|}{MS MARCO Dev} &
        \multicolumn{1}{c}{BEIR} \\
        & MRR@10 & Q Len & D Len & NDCG@10  \\
        \hline
        FLOPs & 0.408  & 208 & 1052   & 0.545 \\
        \hline
        FLOPs + STop & 0.409  & 83  & 364  & 0.540 \\
        \hline
        FLOPs + STop + PTT & 0.408 & 65 & 280 & 0.541 \\
        \hline
        FLOPs + STop + UTT & 0.407  & 46 & 225  & 0.530 \\
        \hline
    \end{tabular}
    }
\end{table}

\subsection{Ablation studies with different sparsification compositions}
Table \ref{tab:effect_topk} further compares different combinations of sparsification techniques.
The first three rows in Table \ref{tab:effect_topk} show the impact of 
adding one sparsification technique incrementally. The result shows that STop can significantly reduce the document and query lengths, 
while retaining competitive relevance (within 1\% degradation), and
 PTT can further 
visibly improve sparsity without compromising the relevance. 
Row 4 replaces PTT with UTT (uniform term thresholding) which
maintains one global threshold for all query terms, and one for all document terms.
FLOPs+STop+UTT 
leads to a shorter query length, but with a near  2\% NDCG drop for BEIR compared to FLOPs+STop+PTT. 

\begin{figure}
   \includegraphics[width=0.8\columnwidth]{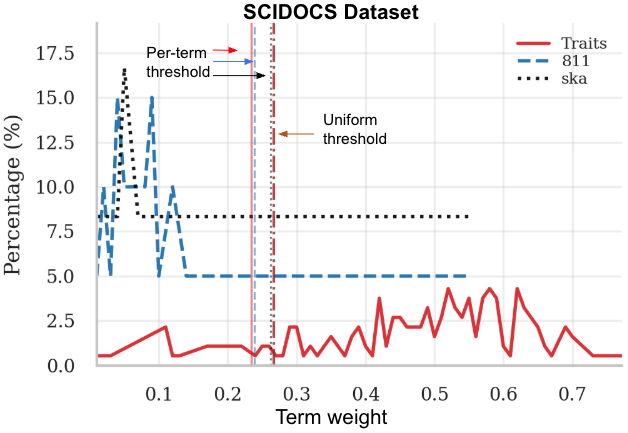}
   \caption{ Weight distribution of three  terms }
   \label{fig:stop_dis}
\end{figure}

Figure~\ref{fig:stop_dis} provides an example that offers insight into the impact of UTT vs PTT applied to
a representation derived by using  FLOPs+STop only on SCIDOCs from the BEIR collection. 
This figure  illustrates the
weight distribution of three representative terms ("Traits", "811", and "ska") from the popular, the medium-popular, and the unpopular tiers,
generated by AdaSparse with FLOPs+STop. 
Each color represents a different term's distribution, and term percentage is based on the frequency of occurrence in the inverted index.
Vertical lines represent one global threshold derived by UTT and three per-term thresholds by PTT. 
Since the curves of weight distributions vary significantly among different terms, a uniform threshold value is unsuitable. 
The global threshold value from UTT is larger than all the PTT thresholds, which results in a small gain in sparsity as 
Table~\ref{tab:effect_topk} suggests, but incurs a greater relevance penalty.

\begin{table}[hbt!]
  \centering
  \caption{Retrieval latency (ms) of Seismic on MS MARCO.}
  \label{tab:pruning_performance_length}
  \resizebox{0.6\columnwidth}{!}{%
  \begin{tabular}{lcccc}
    \toprule
    \multirow{2.5}{*}{Doc Length} & \multicolumn{4}{c}{Query Length} \\
    \cmidrule(lr){2-5}
     & 9 & 17 & 34 & 69 \\
    \midrule
    49    & 1.274 & 1.289 & 1.317 & 1.490 \\
    98    & 1.750 & 1.753 & 1.776 & 1.764 \\
    196   & 2.299 & 2.278 & 2.521 & 2.275 \\
    393   & 3.278 & 3.203 & 3.369 & 3.182 \\
    \bottomrule
  \end{tabular}%
  }
\end{table}
\subsection{Sensitivity of Seismic's retrieval time to query and document lengths}
Compared to the sparsity of SPLADEv3 index~\cite{lassance2024spladev3newbaselinessplade} as shown in Table~\ref{tab:SPLADE-RepLLaMA}
for MS MARCO, Lion-SP-1B's index with AdaSparse is still 2.7x longer in queries 
and 1.6x longer in documents. 
Since the query expansion factor is 
much larger than that of the document, 
we investigate whether retrieval latency scales disproportionately in relation to query length.

Table~\ref{tab:pruning_performance_length} shows the Seismic's latency
growth when we control and vary the average document length of an inverted index
from 49 to 393 and the query length from 9 to 69.  
The index with varying sizes is obtained by truncating query and document vectors
by keeping a fixed number of top weighted terms in a Lion-SP index trained by AdaSparse for MS MARCO. 
There is a significant monotonic increase in latency as document lengths grow, 
whereas the latency remains relatively stable or incurs a modest increase as query length increases.
The above asymmetric latency sensitivity of Seismic retriever to the query and document lengths
matches well with the outcome of AdaSparse in which expansion
is  relatively tight  for documents (3.7x) 
and relatively loose for queries  (8.1x)  for MS MARCO from Table~\ref{tab:adaptive}.

%% file: sigir26concl.tex
\section{Concluding Remarks}

This paper addresses a fundamental computational bottleneck in learned sparse retrieval: the tension between high-dimensional semantic expansion
within a massive  vocabulary space and efficient inference constraints. It  introduces a multi-objective
sparsification scheme with learnable soft top-$K$ regularization and 
per-term thresholding, complementing FLOPs regularization synergistically.

Experimental results with MS MARCO for training and testing, and BEIR datasets for zero-shot retrieval evaluation
indicate that AdaSparse-1B uses 4.2x less space and is 3.8x-4x faster  in Seismic retrieval  than Lion-SP 
, while retaining relevance within 0.74\% difference. 
Compared to Lion-SP-1B with HT, AdaSparse-1B uses 1.8x less storage and is up to 2.1x faster while having 1\% higher relevance.
Compared to MRP, AdaSparse-1B delivers better relevance at a similar sparsity level, especially on BEIR with 2.3\% higher average NDCG@10.
Using LLaMA-3.2 8B significantly boosts MS MARCO Dev performance while the gain for BEIR is modest.

Table~\ref{tab:SPLADE-RepLLaMA} compares Lion-SP/AdaSparse with the latest SPLADE 
version V3~\cite{lassance2024spladev3newbaselinessplade} and RepLLaMA~\cite{ma2024repllama}
to examine
the tradeoff using different language models and between sparse and dense representations.
SPLADEv3 uses a smaller model BERT with 1.7x shorter document length while Lion-SP with AdaSparse gains relevance significantly
on both MS MARCO and BEIR.  
Lion-SP with AdaSparse has 4.3x smaller index size  than RepLLaMA  dense representation  (34GB vs. 145GB) while
RepLLaMA requires much more expensive GPU  hardware for speed.
Relevance of Lion-SP with AdaSparse is better than  RepLLaMA by  1.2\% on Dev,  and by 2.7\% and 3.7\%  on DL19/DL20, 
but has  1.5\% lower average NDCG@10 on BEIR, which suggests room 
for further improvement in the future. 

Note that like SPLADE, Lion-SP and AdaSparse  do not mark titles in  MS MARCO training~\cite{2023SIGIR-Lassance}. 
MRR@10 of  Lion-SP on Dev is strong and AdaSparse retains such a level.
AdaSparse is designed and evaluated for sparsity optimization with Lion-SP.  
Our future work is to apply or extend it to learned sparse representations on different LLMs with a large vocabulary space.
We will also study the use of different sparse retrievers on latency impact, e.g. ~\cite{2026carlson-LSP}.

\vspace{-0.5em}
\begin{table}[htbp]
\begin{scriptsize}
        \centering
        
        \caption{A comparison with  SPLADE and   dense RepLLaMA}
        \vspace{-0.5em}
                \begin{tabular}{l|l |c|c|c|c|c|c }
                \hline
       Methods     & LLM & MM  Dev & DL19 &DL20 &  QLen  & DLen &  BEIR \\ 
		\hline
             SPLADEv3~\cite{lassance2024spladev3newbaselinessplade}  &BERT 110M &  0.402  & 0.723 & 0.754&  24 & 170 & 0.517 \\ 
                \hline
             RepLLaMA dense& LLaMA-2 7B   & 0.412 &0.743 & 0.725  &  4K  & 4K  
&0.551 \\ 
                \hline
             Lion/AdaSparse & LLaMA-3.2 8B& 0.417 &0.763 &0.752 & 85  & 297 
& 0.543 \\ 
                \hline

\end{tabular}
        \label{tab:SPLADE-RepLLaMA}
\end{scriptsize}
\vspace{-0.5em}
\end{table}

\textbf{Acknowledgments.}
We thank anonymous referees for their valuable comments.
This work is supported in part by U.S. NSF IIS-2225942 and a fellowship from Zhu Graduate Support Fund,
and has used the computing resource of the NSF's ACCESS program.
Any opinions, findings, conclusions or recommendations expressed in this material are those of the authors and do not necessarily reflect the views of the U.S. NSF.

%% file: bib/2022extra.bib
@inproceedings{thakur2021beir,
title={{BEIR}: A Heterogeneous Benchmark for Zero-shot Evaluation of Information Retrieval Models},
author={Nandan Thakur and Nils Reimers and Andreas R{\"u}ckl{\'e} and Abhishek Srivastava and Iryna Gurevych},
booktitle={Thirty-fifth Conference on Neural Information Processing Systems Datasets and Benchmarks Track (Round 2)},
year={2021},
url={https://openreview.net/forum?id=wCu6T5xFjeJ}
}


%% file: bib/2022refer.bib
@inproceedings{lassance2022efficiency,
author = {Lassance, Carlos and Clinchant, St\'{e}phane},
title = {An Efficiency Study for SPLADE Models},
year = {2022},
isbn = {9781450387323},
publisher = {Association for Computing Machinery},
address = {New York, NY, USA},
url = {https://doi.org/10.1145/3477495.3531833},
doi = {10.1145/3477495.3531833},
booktitle = {Proceedings of the 45th International ACM SIGIR Conference on Research and Development in Information Retrieval},
pages = {2220–2226},
numpages = {7},
location = {Madrid, Spain},
series = {SIGIR '22}
}


%% file: bib/2024refer.bib
@inproceedings{2024SIGIR-SparseApproximate,
author = {Bruch, Sebastian and Nardini, Franco Maria and Rulli, Cosimo and Venturini, Rossano},
title = {Efficient Inverted Indexes for Approximate Retrieval over Learned Sparse Representations},
year = {2024},
isbn = {9798400704314},
publisher = {Association for Computing Machinery},
address = {New York, NY, USA},
url = {https://doi.org/10.1145/3626772.3657769},
doi = {10.1145/3626772.3657769},
booktitle = {Proceedings of the 47th International ACM SIGIR Conference on Research and Development in Information Retrieval},
pages = {152–162},
numpages = {11},
location = {Washington DC, USA},
series = {SIGIR '24}
}


%% file: bib/2025.bib
@inproceedings{
behnamghader2024llmvec,
title={{LLM}2Vec: Large Language Models Are Secretly Powerful Text Encoders},
author={Parishad BehnamGhader and Vaibhav Adlakha and Marius Mosbach and Dzmitry Bahdanau and Nicolas Chapados and Siva Reddy},
booktitle={First Conference on Language Modeling},
year={2024},
url={https://openreview.net/forum?id=IW1PR7vEBf}
}

@inproceedings{2020NIPS-Lewis-RAG,
author = {Lewis, Patrick and Perez, Ethan and Piktus, Aleksandra and Petroni, Fabio and Karpukhin, Vladimir and Goyal, Naman and K\"{u}ttler, Heinrich and Lewis, Mike and Yih, Wen-tau and Rockt\"{a}schel, Tim and Riedel, Sebastian and Kiela, Douwe},
title = {Retrieval-augmented generation for knowledge-intensive NLP tasks},
year = {2020},
isbn = {9781713829546},
publisher = {Curran Associates Inc.},
address = {Red Hook, NY, USA},
booktitle = {Proceedings of the 34th International Conference on Neural Information Processing Systems},
articleno = {793},
numpages = {16},
location = {Vancouver, BC, Canada},
series = {NIPS '20}
}

@inproceedings{zeng2025scalingsparsedenseretrieval,
author = {Zeng, Hansi and Killingback, Julian and Zamani, Hamed},
title = {Scaling Sparse and Dense Retrieval in Decoder-Only LLMs},
year = {2025},
isbn = {9798400715921},
publisher = {Association for Computing Machinery},
address = {New York, NY, USA},
url = {https://doi.org/10.1145/3726302.3730225},
doi = {10.1145/3726302.3730225},
booktitle = {Proceedings of the 48th International ACM SIGIR Conference on Research and Development in Information Retrieval},
pages = {2679–2684},
numpages = {6},
location = {Padua, Italy},
series = {SIGIR '25}
}

@inproceedings{2025EMNLP-ASE-train,
    title = "Train One Sparse Autoencoder Across Multiple Sparsity Budgets to Preserve Interpretability and Accuracy",
    author = "Balagansky, Nikita  and
      Aksenov, Yaroslav  and
      Laptev, Daniil  and
      Kurochkin, Vadim  and
      Gerasimov, Gleb  and
      Koriagin, Nikita  and
      Gavrilov, Daniil",
    editor = "Christodoulopoulos, Christos  and
      Chakraborty, Tanmoy  and
      Rose, Carolyn  and
      Peng, Violet",
    booktitle = "Proceedings of the 2025 Conference on Empirical Methods in Natural Language Processing",
    month = nov,
    year = "2025",
    address = "Suzhou, China",
    publisher = "Association for Computational Linguistics",
    url = "https://aclanthology.org/2025.emnlp-main.515/",
    doi = "10.18653/v1/2025.emnlp-main.515",
    pages = "10182--10190",
    ISBN = "979-8-89176-332-6"
}

@inproceedings{2025EMNLP-SAE-dense-Park,
    title = "Decoding Dense Embeddings: Sparse Autoencoders for Interpreting and Discretizing Dense Retrieval",
    author = "Park, Seongwan  and
      Kim, Taeklim  and
      Ko, Youngjoong",
    editor = "Christodoulopoulos, Christos  and
      Chakraborty, Tanmoy  and
      Rose, Carolyn  and
      Peng, Violet",
    booktitle = "Proceedings of the 2025 Conference on Empirical Methods in Natural Language Processing",
    month = nov,
    year = "2025",
    address = "Suzhou, China",
    publisher = "Association for Computational Linguistics",
    url = "https://aclanthology.org/2025.emnlp-main.1345/",
    doi = "10.18653/v1/2025.emnlp-main.1345",
    pages = "26479--26496",
    ISBN = "979-8-89176-332-6"
}

@inproceedings{
2024ICLR-SAE-Huben,
title={Sparse Autoencoders Find Highly Interpretable Features in Language Models},
author={Robert Huben and Hoagy Cunningham and Logan Riggs Smith and Aidan Ewart and Lee Sharkey},
booktitle={The Twelfth International Conference on Learning Representations},
year={2024},
url={https://openreview.net/forum?id=F76bwRSLeK}
}

@article{bricken2023monosemanticity,
       title={Towards Monosemanticity: Decomposing Language Models With Dictionary Learning},
       author={Bricken, Trenton and Templeton, Adly and Batson, Joshua and Chen, Brian and Jermyn, Adam and Conerly, Tom and Turner, Nick and Anil, Cem and Denison, Carson and Askell, Amanda and Lasenby, Robert and Wu, Yifan and Kravec, Shauna and Schiefer, Nicholas and Maxwell, Tim and Joseph, Nicholas and Hatfield-Dodds, Zac and Tamkin, Alex and Nguyen, Karina and McLean, Brayden and Burke, Josiah E and Hume, Tristan and Carter, Shan and Henighan, Tom and Olah, Christopher},
       year={2023},
       journal={Transformer Circuits Thread},
       note={https://transformer-circuits.pub/2023/monosemantic-features/index.html}
    }


%% file: bib/2026wentai.bib
@inproceedings{2023ICTIR-CSurF-CMU,
author = {Fan, Zhen and Gao, Luyu and Callan, Jamie},
title = {CSurF: Sparse Lexical Retrieval through Contextualized Surface Forms},
year = {2023},
isbn = {9798400700736},
publisher = {Association for Computing Machinery},
address = {New York, NY, USA},
url = {https://doi.org/10.1145/3578337.3605126},
doi = {10.1145/3578337.3605126},
booktitle = {Proceedings of the 2023 ACM SIGIR International Conference on Theory of Information Retrieval},
pages = {65–75},
numpages = {11},
location = {Taipei, Taiwan},
series = {ICTIR '23}
}

@misc{bajaj2018msmarcohumangenerated,
      title={MS MARCO: A Human Generated MAchine Reading COmprehension Dataset}, 
      author={Payal Bajaj and Daniel Campos and Nick Craswell and Li Deng and Jianfeng Gao and Xiaodong Liu and Rangan Majumder and Andrew McNamara and Bhaskar Mitra and Tri Nguyen and Mir Rosenberg and Xia Song and Alina Stoica and Saurabh Tiwary and Tong Wang},
      year={2018},
      eprint={1611.09268},
      archivePrefix={arXiv},
      primaryClass={cs.CL},
      url={https://arxiv.org/abs/1611.09268}, 
}

@misc{zhou2024retrievalbaseddisentangledrepresentationlearning,
      title={Retrieval-based Disentangled Representation Learning with Natural Language Supervision}, 
      author={Jiawei Zhou and Xiaoguang Li and Lifeng Shang and Xin Jiang and Qun Liu and Lei Chen},
      year={2024},
      eprint={2212.07699},
      archivePrefix={arXiv},
      primaryClass={cs.CL},
      url={https://arxiv.org/abs/2212.07699}, 
}

@misc{reif2025vocabdietreshapingvocabulary,
      title={Vocab Diet: Reshaping the Vocabulary of LLMs with Vector Arithmetic}, 
      author={Yuval Reif and Guy Kaplan and Roy Schwartz},
      year={2025},
      eprint={2510.17001},
      archivePrefix={arXiv},
      primaryClass={cs.CL},
      url={https://arxiv.org/abs/2510.17001}, 
}

@misc{lassance2024spladev3newbaselinessplade,
      title={SPLADE-v3: New baselines for SPLADE}, 
      author={Carlos Lassance and Hervé Déjean and Thibault Formal and Stéphane Clinchant},
      year={2024},
      eprint={2403.06789},
      archivePrefix={arXiv},
      primaryClass={cs.IR},
      url={https://arxiv.org/abs/2403.06789}, 
}

@misc{grattafiori2024llama3herdmodels,
      title={The Llama 3 Herd of Models}, 
      author={Aaron Grattafiori and Abhimanyu Dubey and Abhinav Jauhri and Abhinav Pandey and Abhishek Kadian and Ahmad Al-Dahle and Aiesha Letman and Akhil Mathur and Alan Schelten and Alex Vaughan and Amy Yang and Angela Fan and Anirudh Goyal and Anthony Hartshorn and Aobo Yang and Archi Mitra and Archie Sravankumar and Artem Korenev and Arthur Hinsvark and Arun Rao and Aston Zhang and Aurelien Rodriguez and Austen Gregerson and Ava Spataru and Baptiste Roziere and Bethany Biron and Binh Tang and Bobbie Chern and Charlotte Caucheteux and Chaya Nayak and Chloe Bi and Chris Marra and Chris McConnell and Christian Keller and Christophe Touret and Chunyang Wu and Corinne Wong and Cristian Canton Ferrer and Cyrus Nikolaidis and Damien Allonsius and Daniel Song and Danielle Pintz and Danny Livshits and Danny Wyatt and David Esiobu and Dhruv Choudhary and Dhruv Mahajan and Diego Garcia-Olano and Diego Perino and Dieuwke Hupkes and Egor Lakomkin and Ehab AlBadawy and Elina Lobanova and Emily Dinan and Eric Michael Smith and Filip Radenovic and Francisco Guzmán and Frank Zhang and Gabriel Synnaeve and Gabrielle Lee and Georgia Lewis Anderson and Govind Thattai and Graeme Nail and Gregoire Mialon and Guan Pang and Guillem Cucurell and Hailey Nguyen and Hannah Korevaar and Hu Xu and Hugo Touvron and Iliyan Zarov and Imanol Arrieta Ibarra and Isabel Kloumann and Ishan Misra and Ivan Evtimov and Jack Zhang and Jade Copet and Jaewon Lee and Jan Geffert and Jana Vranes and Jason Park and Jay Mahadeokar and Jeet Shah and Jelmer van der Linde and Jennifer Billock and Jenny Hong and Jenya Lee and Jeremy Fu and Jianfeng Chi and Jianyu Huang and Jiawen Liu and Jie Wang and Jiecao Yu and Joanna Bitton and Joe Spisak and Jongsoo Park and Joseph Rocca and Joshua Johnstun and Joshua Saxe and Junteng Jia and Kalyan Vasuden Alwala and Karthik Prasad and Kartikeya Upasani and Kate Plawiak and Ke Li and Kenneth Heafield and Kevin Stone and Khalid El-Arini and Krithika Iyer and Kshitiz Malik and Kuenley Chiu and Kunal Bhalla and Kushal Lakhotia and Lauren Rantala-Yeary and Laurens van der Maaten and Lawrence Chen and Liang Tan and Liz Jenkins and Louis Martin and Lovish Madaan and Lubo Malo and Lukas Blecher and Lukas Landzaat and Luke de Oliveira and Madeline Muzzi and Mahesh Pasupuleti and Mannat Singh and Manohar Paluri and Marcin Kardas and Maria Tsimpoukelli and Mathew Oldham and Mathieu Rita and Maya Pavlova and Melanie Kambadur and Mike Lewis and Min Si and Mitesh Kumar Singh and Mona Hassan and Naman Goyal and Narjes Torabi and Nikolay Bashlykov and Nikolay Bogoychev and Niladri Chatterji and Ning Zhang and Olivier Duchenne and Onur Çelebi and Patrick Alrassy and Pengchuan Zhang and Pengwei Li and Petar Vasic and Peter Weng and Prajjwal Bhargava and Pratik Dubal and Praveen Krishnan and Punit Singh Koura and Puxin Xu and Qing He and Qingxiao Dong and Ragavan Srinivasan and Raj Ganapathy and Ramon Calderer and Ricardo Silveira Cabral and Robert Stojnic and Roberta Raileanu and Rohan Maheswari and Rohit Girdhar and Rohit Patel and Romain Sauvestre and Ronnie Polidoro and Roshan Sumbaly and Ross Taylor and Ruan Silva and Rui Hou and Rui Wang and Saghar Hosseini and Sahana Chennabasappa and Sanjay Singh and Sean Bell and Seohyun Sonia Kim and Sergey Edunov and Shaoliang Nie and Sharan Narang and Sharath Raparthy and Sheng Shen and Shengye Wan and Shruti Bhosale and Shun Zhang and Simon Vandenhende and Soumya Batra and Spencer Whitman and Sten Sootla and Stephane Collot and Suchin Gururangan and Sydney Borodinsky and Tamar Herman and Tara Fowler and Tarek Sheasha and Thomas Georgiou and Thomas Scialom and Tobias Speckbacher and Todor Mihaylov and Tong Xiao and Ujjwal Karn and Vedanuj Goswami and Vibhor Gupta and Vignesh Ramanathan and Viktor Kerkez and Vincent Gonguet and Virginie Do and Vish Vogeti and Vítor Albiero and Vladan Petrovic and Weiwei Chu and Wenhan Xiong and Wenyin Fu and Whitney Meers and Xavier Martinet and Xiaodong Wang and Xiaofang Wang and Xiaoqing Ellen Tan and Xide Xia and Xinfeng Xie and Xuchao Jia and Xuewei Wang and Yaelle Goldschlag and Yashesh Gaur and Yasmine Babaei and Yi Wen and Yiwen Song and Yuchen Zhang and Yue Li and Yuning Mao and Zacharie Delpierre Coudert and Zheng Yan and Zhengxing Chen and Zoe Papakipos and Aaditya Singh and Aayushi Srivastava and Abha Jain and Adam Kelsey and Adam Shajnfeld and Adithya Gangidi and Adolfo Victoria and Ahuva Goldstand and Ajay Menon and Ajay Sharma and Alex Boesenberg and Alexei Baevski and Allie Feinstein and Amanda Kallet and Amit Sangani and Amos Teo and Anam Yunus and Andrei Lupu and Andres Alvarado and Andrew Caples and Andrew Gu and Andrew Ho and Andrew Poulton and Andrew Ryan and Ankit Ramchandani and Annie Dong and Annie Franco and Anuj Goyal and Aparajita Saraf and Arkabandhu Chowdhury and Ashley Gabriel and Ashwin Bharambe and Assaf Eisenman and Azadeh Yazdan and Beau James and Ben Maurer and Benjamin Leonhardi and Bernie Huang and Beth Loyd and Beto De Paola and Bhargavi Paranjape and Bing Liu and Bo Wu and Boyu Ni and Braden Hancock and Bram Wasti and Brandon Spence and Brani Stojkovic and Brian Gamido and Britt Montalvo and Carl Parker and Carly Burton and Catalina Mejia and Ce Liu and Changhan Wang and Changkyu Kim and Chao Zhou and Chester Hu and Ching-Hsiang Chu and Chris Cai and Chris Tindal and Christoph Feichtenhofer and Cynthia Gao and Damon Civin and Dana Beaty and Daniel Kreymer and Daniel Li and David Adkins and David Xu and Davide Testuggine and Delia David and Devi Parikh and Diana Liskovich and Didem Foss and Dingkang Wang and Duc Le and Dustin Holland and Edward Dowling and Eissa Jamil and Elaine Montgomery and Eleonora Presani and Emily Hahn and Emily Wood and Eric-Tuan Le and Erik Brinkman and Esteban Arcaute and Evan Dunbar and Evan Smothers and Fei Sun and Felix Kreuk and Feng Tian and Filippos Kokkinos and Firat Ozgenel and Francesco Caggioni and Frank Kanayet and Frank Seide and Gabriela Medina Florez and Gabriella Schwarz and Gada Badeer and Georgia Swee and Gil Halpern and Grant Herman and Grigory Sizov and Guangyi and Zhang and Guna Lakshminarayanan and Hakan Inan and Hamid Shojanazeri and Han Zou and Hannah Wang and Hanwen Zha and Haroun Habeeb and Harrison Rudolph and Helen Suk and Henry Aspegren and Hunter Goldman and Hongyuan Zhan and Ibrahim Damlaj and Igor Molybog and Igor Tufanov and Ilias Leontiadis and Irina-Elena Veliche and Itai Gat and Jake Weissman and James Geboski and James Kohli and Janice Lam and Japhet Asher and Jean-Baptiste Gaya and Jeff Marcus and Jeff Tang and Jennifer Chan and Jenny Zhen and Jeremy Reizenstein and Jeremy Teboul and Jessica Zhong and Jian Jin and Jingyi Yang and Joe Cummings and Jon Carvill and Jon Shepard and Jonathan McPhie and Jonathan Torres and Josh Ginsburg and Junjie Wang and Kai Wu and Kam Hou U and Karan Saxena and Kartikay Khandelwal and Katayoun Zand and Kathy Matosich and Kaushik Veeraraghavan and Kelly Michelena and Keqian Li and Kiran Jagadeesh and Kun Huang and Kunal Chawla and Kyle Huang and Lailin Chen and Lakshya Garg and Lavender A and Leandro Silva and Lee Bell and Lei Zhang and Liangpeng Guo and Licheng Yu and Liron Moshkovich and Luca Wehrstedt and Madian Khabsa and Manav Avalani and Manish Bhatt and Martynas Mankus and Matan Hasson and Matthew Lennie and Matthias Reso and Maxim Groshev and Maxim Naumov and Maya Lathi and Meghan Keneally and Miao Liu and Michael L. Seltzer and Michal Valko and Michelle Restrepo and Mihir Patel and Mik Vyatskov and Mikayel Samvelyan and Mike Clark and Mike Macey and Mike Wang and Miquel Jubert Hermoso and Mo Metanat and Mohammad Rastegari and Munish Bansal and Nandhini Santhanam and Natascha Parks and Natasha White and Navyata Bawa and Nayan Singhal and Nick Egebo and Nicolas Usunier and Nikhil Mehta and Nikolay Pavlovich Laptev and Ning Dong and Norman Cheng and Oleg Chernoguz and Olivia Hart and Omkar Salpekar and Ozlem Kalinli and Parkin Kent and Parth Parekh and Paul Saab and Pavan Balaji and Pedro Rittner and Philip Bontrager and Pierre Roux and Piotr Dollar and Polina Zvyagina and Prashant Ratanchandani and Pritish Yuvraj and Qian Liang and Rachad Alao and Rachel Rodriguez and Rafi Ayub and Raghotham Murthy and Raghu Nayani and Rahul Mitra and Rangaprabhu Parthasarathy and Raymond Li and Rebekkah Hogan and Robin Battey and Rocky Wang and Russ Howes and Ruty Rinott and Sachin Mehta and Sachin Siby and Sai Jayesh Bondu and Samyak Datta and Sara Chugh and Sara Hunt and Sargun Dhillon and Sasha Sidorov and Satadru Pan and Saurabh Mahajan and Saurabh Verma and Seiji Yamamoto and Sharadh Ramaswamy and Shaun Lindsay and Shaun Lindsay and Sheng Feng and Shenghao Lin and Shengxin Cindy Zha and Shishir Patil and Shiva Shankar and Shuqiang Zhang and Shuqiang Zhang and Sinong Wang and Sneha Agarwal and Soji Sajuyigbe and Soumith Chintala and Stephanie Max and Stephen Chen and Steve Kehoe and Steve Satterfield and Sudarshan Govindaprasad and Sumit Gupta and Summer Deng and Sungmin Cho and Sunny Virk and Suraj Subramanian and Sy Choudhury and Sydney Goldman and Tal Remez and Tamar Glaser and Tamara Best and Thilo Koehler and Thomas Robinson and Tianhe Li and Tianjun Zhang and Tim Matthews and Timothy Chou and Tzook Shaked and Varun Vontimitta and Victoria Ajayi and Victoria Montanez and Vijai Mohan and Vinay Satish Kumar and Vishal Mangla and Vlad Ionescu and Vlad Poenaru and Vlad Tiberiu Mihailescu and Vladimir Ivanov and Wei Li and Wenchen Wang and Wenwen Jiang and Wes Bouaziz and Will Constable and Xiaocheng Tang and Xiaojian Wu and Xiaolan Wang and Xilun Wu and Xinbo Gao and Yaniv Kleinman and Yanjun Chen and Ye Hu and Ye Jia and Ye Qi and Yenda Li and Yilin Zhang and Ying Zhang and Yossi Adi and Youngjin Nam and Yu and Wang and Yu Zhao and Yuchen Hao and Yundi Qian and Yunlu Li and Yuzi He and Zach Rait and Zachary DeVito and Zef Rosnbrick and Zhaoduo Wen and Zhenyu Yang and Zhiwei Zhao and Zhiyu Ma},
      year={2024},
      eprint={2407.21783},
      archivePrefix={arXiv},
      primaryClass={cs.AI},
      url={https://arxiv.org/abs/2407.21783}, 
}

@misc{2026arXivSINDI,
      title={SINDI: an Efficient Index for Approximate Maximum Inner Product Search on Sparse Vectors}, 
      author={Ruoxuan Li and Xiaoyao Zhong and Jiabao Jin and Peng Cheng and Wangze Ni and Zhitao Shen and Wei Jia and Xiangyu Wang and Heng Tao Shen and Jingkuan Song},
      year={2026},
      eprint={2509.08395},
      archivePrefix={ {ICDE 2026 and arXiv} },
      primaryClass={cs.DB},
      url={https://arxiv.org/abs/2509.08395}, 
}

@inproceedings{2026carlson-LSP,
    title={Efficient Sparse Retrieval with Lightweight Superblock Pruning},
    author={Parker Carlson and Wentai Xie and Rohil Shah and Tao Yang},
    booktitle = "Proceedings of the 49th International ACM SIGIR Conference on Research and Development in Information Retrieval (SIGIR ’26)",
    month = jul,
    year = "2026",
    address = "New York, NY, USA",
    publisher = "Association for Computing Machinery",
    doi = "10.1145/3805712.3809739", 
    ISBN = "979-8-4007-2599-9/2026/07",
}


%% file: bib/jinjin_thesis.bib
@inproceedings{Mallia2017VBMW,
author = {Mallia, Antonio and Ottaviano, Giuseppe and Porciani, Elia and Tonellotto, Nicola and Venturini, Rossano},
title = {Faster BlockMax WAND with Variable-sized Blocks},
year = {2017},
isbn = {9781450350228},
publisher = {Association for Computing Machinery},
address = {New York, NY, USA},
url = {https://doi.org/10.1145/3077136.3080780},
doi = {10.1145/3077136.3080780},
booktitle = {Proceedings of the 40th International ACM SIGIR Conference on Research and Development in Information Retrieval},
pages = {625–634},
numpages = {10},
location = {Shinjuku, Tokyo, Japan},
series = {SIGIR '17}
}


%% file: bib/mise.bib
@article{Craswell2020OverviewOT,
  title={Overview of the TREC 2020 Deep Learning Track},
  author={Nick Craswell and Bhaskar Mitra and Emine Yilmaz and Daniel Fernando Campos and Ellen M. Voorhees},
  journal={ArXiv},
  year={2020},
  volume={abs/2102.07662}
}


%% file: bib/ranking.bib
@article{Dai2020deepct,
  title={Context-Aware Term Weighting For First Stage Passage Retrieval},
  author={Zhuyun Dai and Jamie Callan},
  journal={SIGIR},
  year={2020}
}

@inproceedings{Mallia2021deepimpact,
author = {Mallia, Antonio and Khattab, Omar and Suel, Torsten and Tonellotto, Nicola},
title = {Learning Passage Impacts for Inverted Indexes},
year = {2021},
isbn = {9781450380379},
publisher = {Association for Computing Machinery},
address = {New York, NY, USA},
url = {https://doi.org/10.1145/3404835.3463030},
doi = {10.1145/3404835.3463030},
booktitle = {Proceedings of the 44th International ACM SIGIR Conference on Research and Development in Information Retrieval},
pages = {1723–1727},
numpages = {5},
location = {Virtual Event, Canada},
series = {SIGIR '21}
}

@article{Lin2021unicoil,
  author       = {Jimmy Lin and
                  Xueguang Ma},
  title        = {A Few Brief Notes on DeepImpact, COIL, and a Conceptual Framework
                  for Information Retrieval Techniques},
  journal      = {CoRR},
  volume       = {abs/2106.14807},
  year         = {2021},
  url          = {https://arxiv.org/abs/2106.14807},
  eprinttype    = {arXiv},
  eprint       = {2106.14807},
  bibsource    = {dblp computer science bibliography, https://dblp.org}
}

@article{Hofsttter2020marginMSE,
  title={Improving Efficient Neural Ranking Models with Cross-Architecture Knowledge Distillation},
  author={Sebastian Hofst{\"a}tter and Sophia Althammer and Michael Schr{\"o}der and Mete Sertkan and Allan Hanbury},
  journal={ArXiv},
  year={2020},
  volume={abs/2010.02666}
}

@article{2021NAACL-Gao-COIL,
  author    = {Luyu Gao and
               Zhuyun Dai and
               Jamie Callan},
  title     = {{COIL:} Revisit Exact Lexical Match in Information Retrieval with
               Contextualized Inverted List},
  journal   = {NAACL},
  year      = {2021},
}

@article{Formal2021SPLADEV2,
  title={{SPLADE} v2: Sparse Lexical and Expansion Model for Information Retrieval},
  author={Thibault Formal and C. Lassance and Benjamin Piwowarski and St{\'e}phane Clinchant},
  journal={ArXiv},
  year={2021},
  volume={abs/2109.10086}
}

@inproceedings{Formal2021SPLADE,
author = {Formal, Thibault and Piwowarski, Benjamin and Clinchant, St\'{e}phane},
title = {SPLADE: Sparse Lexical and Expansion Model for First Stage Ranking},
year = {2021},
isbn = {9781450380379},
publisher = {Association for Computing Machinery},
address = {New York, NY, USA},
url = {https://doi.org/10.1145/3404835.3463098},
doi = {10.1145/3404835.3463098},
booktitle = {Proceedings of the 44th International ACM SIGIR Conference on Research and Development in Information Retrieval},
pages = {2288–2292},
numpages = {5},
location = {Virtual Event, Canada},
series = {SIGIR '21}
}

@inproceedings{2023SIGIR-Qiao,
author = {Qiao, Yifan and Yang, Yingrui and He, Shanxiu and Yang, Tao},
title = {Representation Sparsification with Hybrid Thresholding for Fast SPLADE-based Document Retrieval},
year = {2023},
isbn = {9781450394086},
publisher = {Association for Computing Machinery},
address = {New York, NY, USA},
url = {https://doi.org/10.1145/3539618.3592051},
doi = {10.1145/3539618.3592051},
booktitle = {Proceedings of the 46th International ACM SIGIR Conference on Research and Development in Information Retrieval},
pages = {2329–2333},
numpages = {5},
location = {Taipei, Taiwan},
series = {SIGIR '23}
}

@inproceedings{2023SIGIR-SPLADE-pruning,
author = {Lassance, Carlos and Lupart, Simon and D\'{e}jean, Herv\'{e} and Clinchant, St\'{e}phane and Tonellotto, Nicola},
title = {A Static Pruning Study on Sparse Neural Retrievers},
year = {2023},
isbn = {9781450394086},
publisher = {Association for Computing Machinery},
address = {New York, NY, USA},
booktitle = {Proc. of the 46th International ACM SIGIR Conference on Research and Development in Information Retrieval},
pages = {1771–1775},
numpages = {5},
location = {Taipei, Taiwan},
series = {SIGIR '23}
}

@inproceedings{shen2023lexmae,
  author       = {Tao Shen and
                  Xiubo Geng and
                  Chongyang Tao and
                  Can Xu and
                  Xiaolong Huang and
                  Binxing Jiao and
                  Linjun Yang and
                  Daxin Jiang},
  title        = {LexMAE: Lexicon-Bottlenecked Pretraining for Large-Scale Retrieval},
  booktitle    = {The Eleventh International Conference on Learning Representations,
                  {ICLR} 2023, Kigali, Rwanda, May 1-5, 2023},
  publisher    = {International Conference on Learning Representations (ICLR)},
  year         = {2023},
  url          = {https://openreview.net/forum?id=PfpEtB3-csK},
  bibsource    = {dblp computer science bibliography, https://dblp.org}
}

@inproceedings{2023SIGIR-Lassance,
author = {Lassance, Carlos and Clinchant, Stephane},
title = {The Tale of Two MSMARCO - and Their Unfair Comparisons},
year = {2023}, 
publisher = {ACM},
address = {New York, NY, USA},
booktitle = {Proceed. of the 46th International ACM SIGIR Conference on Research and Development in Information Retrieval},
pages = {2431–2435},
numpages = {5},
series = {SIGIR '23}
}


%% file: bib/reference.bib
@article{Zamani2018SNRM,
  title={From Neural Re-Ranking to Neural Ranking: Learning a Sparse Representation for Inverted Indexing},
  author={Hamed Zamani and Mostafa Dehghani and William Bruce Croft and Erik G. Learned-Miller and J. Kamps},
  journal={Proceedings of the 27th ACM International Conference on Information and Knowledge Management},
  year={2018}
}

@inproceedings{Formal_etal_SIGIR2022_splade++,
author = {Formal, Thibault and Lassance, Carlos and Piwowarski, Benjamin and Clinchant, St\'{e}phane},
title = {From Distillation to Hard Negative Sampling: Making Sparse Neural IR Models More Effective},
year = {2022},
isbn = {9781450387323},
publisher = {Association for Computing Machinery},
address = {New York, NY, USA},
url = {https://doi.org/10.1145/3477495.3531857},
doi = {10.1145/3477495.3531857},
booktitle = {Proceedings of the 45th International ACM SIGIR Conference on Research and Development in Information Retrieval},
pages = {2353–2359},
numpages = {7},
location = {Madrid, Spain},
series = {SIGIR '22}
}

@inproceedings{ma2024repllama,
author = {Ma, Xueguang and Wang, Liang and Yang, Nan and Wei, Furu and Lin, Jimmy},
title = {Fine-Tuning LLaMA for Multi-Stage Text Retrieval},
year = {2024},
isbn = {9798400704314},
publisher = {Association for Computing Machinery},
address = {New York, NY, USA},
url = {https://doi.org/10.1145/3626772.3657951},
doi = {10.1145/3626772.3657951},
booktitle = {Proceedings of the 47th International ACM SIGIR Conference on Research and Development in Information Retrieval},
pages = {2421–2425},
numpages = {5},
location = {Washington DC, USA},
series = {SIGIR '24}
}


%% file: bib/thres.bib
@article{2021-Yang-Masking, 
  author    = {Jheng{-}Hong Yang and
               Xueguang Ma and
               Jimmy Lin},
  title     = {Sparsifying Sparse Representations for Passage Retrieval by Top-k
               Masking},
  journal   = {CoRR},
  volume    = {abs/2112.09628},
  year      = {2021},
  url       = {https://arxiv.org/abs/2112.09628},
  eprinttype = {arXiv},
  eprint    = {2112.09628},
  bibsource = {dblp computer science bibliography, https://dblp.org}
}

@inproceedings{
Paria2020Minimizing,
title={Minimizing FLOPs to Learn Efficient Sparse Representations},
author={Biswajit Paria and Chih-Kuan Yeh and Ian E.H. Yen and Ning Xu and Pradeep Ravikumar and Barnabás Póczos},
booktitle={International Conference on Learning Representations},
year={2020},
url={https://openreview.net/forum?id=SygpC6Ntvr}
}

@inproceedings{2001SIGIR-Carmel,
author = {Carmel, David and Cohen, Doron and Fagin, Ronald and Farchi, Eitan and Herscovici, Michael and Maarek, Yoelle S. and Soffer, Aya},
title = {Static Index Pruning for Information Retrieval Systems},
year = {2001},
isbn = {1581133316},
publisher = {Association for Computing Machinery},
address = {New York, NY, USA},
url = {https://doi.org/10.1145/383952.383958},
doi = {10.1145/383952.383958},
booktitle = {Proc. of SIGIR},
pages = {43–50},
numpages = {8},
location = {New Orleans, Louisiana, USA},
series = {SIGIR '01}
}

@inproceedings{2007SIGIR-Blanco,
author = {Blanco, Roi and Barreiro, Alvaro},
title = {Boosting Static Pruning of Inverted Files},
year = {2007},
isbn = {9781595935977},
publisher = {Association for Computing Machinery},
address = {New York, NY, USA},
url = {https://doi.org/10.1145/1277741.1277904},
doi = {10.1145/1277741.1277904},
booktitle = {Proc. of SIGIR},
pages = {777–778},
numpages = {2},
location = {Amsterdam, The Netherlands},
series = {SIGIR '07}
}

@inproceedings{2006CIKM-Buttcher,
author = {B\"{u}ttcher, Stefan and Clarke, Charles L. A.},
title = {A Document-Centric Approach to Static Index Pruning in Text Retrieval Systems},
year = {2006},
isbn = {1595934332},
publisher = {Association for Computing Machinery},
address = {New York, NY, USA},
url = {https://doi.org/10.1145/1183614.1183644},
doi = {10.1145/1183614.1183644},
booktitle = {Proceedings of the 15th ACM International Conference on Information and Knowledge Management},
pages = {182–189},
numpages = {8},
location = {Arlington, Virginia, USA},
series = {CIKM '06}
}


%% file: bib/url.bib
@article{lionRepo,
author = "LION",
title = "https://github.com/HansiZeng/scaling-retriever",
year = "2025"
}
